# Simulations of Fibre Orientation in Dilute Suspensions with Front Moving in the Filling Process of a Rectangular Channel Using Level Set Method

Hua-Shu Dou[1*], Boo Cheong Khoo[2] Nhan Phan-Thien[2],
Khoon Seng Yeo[2], Rong Zheng[3]

[1]Temasek Laboratories
National University of Singapore,  Singapore 117508
[2]Department of Mechanical Engineering
National University of Singapore, Singapore 119260
[3]Moldflow Pty. Ltd, 259-261 Colchester Road, Kilsyth, Vic 3137, Australia
*tsldh@nus.edu.sg; huashudou@yahoo.com

**Abstract**   The simulation of fibre orientation in dilute suspension with a front moving is carried out using the projection and level set methods. The motion of fibres is described using the Jeffery equation and the contribution of fibres to the flow is accounted for by the configuration field method. The dilute suspension of short fibres in Newtonian fluids is considered. The governing Navier-Stokes equation for the fluid flow is solved using the projection method with finite difference scheme, while the fibre-related equations are directly solved with the Runge-Kutta method. In the present study for fibres in dilute suspension flow as for injection molding, the effects of various flow and material parameters on the fibre orientation and the velocity distributions as well as the shapes of the leading flow front are found and discussed. Our findings indicate that the presence of fibres motion has little influence on the front shape in the ranges of fibre parameters studied at the fixed Reynolds number. Influence of changing fibre parameters only causes variation of front shape in the region near the wall and the front shape in the central core area does not vary much with the fibre parameters. On the other hand, the fibre motion has strong influence on the distributions of the streamwise and transverse velocities in the fountain flow. Fibre motion produces strong normal stress near the wall which leads to the reduction of transversal velocity as compared to the Newtonian flow without fibres and in turn the streamwise velocity near the wall is increased. Thus, the fibre addition to the flow weakens the strength of the fountain flow. The Reynolds number has also displayed significant influence on the distribution of the streamwise velocity behind the flow front for a given fibre concentration.  It is also found that the fibre orientation is not always along the direction of velocity vector in the process of mold filling. In the region of the fountain flow, the fibre near the centerline is more oriented cross the streamwise direction comparing to that in the region far behind the flow front. This leads to that the fibre near the centreline in the region of fountain flow is more extended along the transverse direction. Since fibre orientation in the suspension flow and the shape of the flow front have great bearing on the quality of the product made from injection molding, this study has much implications for engineering applications. These results can also be useful in other field dealing with fibre suspensions.

## 1. Introduction

Simulation of fibre suspensions as in the injecting moulding process can lead to improvement in product quality and characteristics. The quality or accuracy of such simulation of fibre suspension depends largely on the appropriate description of the constitutive equations relating the fibre properties in the flows and the correct prediction of property of the fibre flows. In the past few decades or so, several models have been proposed for dilute suspensions and various numerical methods have been developed to a large extent (Hinch and Leal 1976; Dinh and Amstrong 1984; Advani and Tucker 1987; Shaqfeh and Fredrickson 1990; Phan-Thien and Graham 1991). The reliability of the numerical simulation depends to a large extent on the modelling of the constitutive behaviours. The rheological modelling employed in the experiments and simulations has been reviewed in Petrie (1999).

Generally, most simulations reported in the literature use the continuous mechanics approach to solve the governing Navier-Stokes equations for the flow and in which the contribution of the fibre motion is included in the various stress terms. There are three main numerical techniques accounting for such said stress contributions from fibre suspension. One is based on the assumption that the fibre is in full-alignment (along) with the velocity vector (Lipscomb et al. 1988; Chiba and Nakamura 1990; Rosenberg and Denn 1990; Baloch and Webster 1995). The second solves for the stress tensor in the flow field. In this method, the fibre stress is modelled by a constitutive equation of the fibre orientation tensor (Advani and Tucker 1987; Reddy and Mitchell 2001; VerWeyst and Tucker 2002). The third is that the orientation tensor of fibre is modelled by Brownian configuration field method (Hulsen et al. 1997; Fan et al. 1999; Chiba et al. 2001). Azaiez and Guenette (2002) compared the fibre aligned-assumption and the solution of orientation tensors which is defined as dyadic products of the orientation vector through their FEM solution for the benchmark problem of 4:1 contraction duct. Their results show that the coupling between the flow and the fibre orientation is very important in the modelling. Chinesta and Poitou (2002) also showed that the coupling between the flow kinematics and the fibre orientation in dilute fibre suspensions is necessary for the accurate simulation of the flow behaviour, which is of industrial interests for the process optimization to align the fibres in structural pieces in the main stress directions.

The fibre suspension is generally classified as dilute, semi-dilute and concentrated as follows. A suspension of uniform distributed fibres can be characterized by its volume fraction $\phi$ and the fibre aspect ratio $a_r$ (the ratio of the fibre length $l$ to the fibre diameter $d$). If $\phi a_r^2 < 1$, the suspension is said to be dilute. For a dilute suspension, each fibre can move freely and there is no interaction between fibres. Hence the volume-averaged stress can be obtained by Jeffery's solution (Jeffery 1922). If $1 < \phi a_r^2 < a_r$ and $1 < \phi a_r$, the suspension is said to be semi-dilute and concentrated, respectively. For these latter two cases, the hydrodynamic interactions among fibres affect the motion of the fibre. Thus, the fibre orientation is strongly controlled by the anisotropy of fibre motions. Folgar and Tucker (1984) proposed a model, in which the fibre interactions, modeled as a diffusion process, are added into the Jeffery's equation. Fan et al. (1999) have successfully simulated the flows of dilute and semi-dilute suspensions around a sphere in a channel based on this model.

The level set method has been developed and used in the tracking of the flow front evolution, and has been demonstrated to be very useful with great ease of implementation (Sethian and Smereka 2003; Sussman et al. 1994; Sussman et al. 1998). Thus far, it has not been applied to the fibre suspension flows, although it was employed recently for simulating the film flows of shear thinning flows on an inclined plate (Dou et al. 2004). In this said method, a level set function for capturing the interface between two fluids is combined with a variable density projection method to allow for computation of a two-phase flow where the interface can merge and break. A smoothed distance function formulation of the level set method at the sharp interface enables one to compute flows with large density ratios (1000/1) and flows that are surface tension driven.

In this paper, the methodology for simulating the front evolution of fibre suspension flow is proposed using both the projection and level set methods. The dilute suspensions of short fibres in Newtonian fluids are considered. The Brownian configuration field method in literature (Fan et al. 1999; Hulsen et al 1997; Ottinger et al. 1997) is employed to solve the fibre motion and to account for the fibre contributions to the various stress terms in the flow equations associated with the governing Navier-Stokes equation. The objective of the study is to investigate the effect of fibre behaviour on the front shape and velocity field as well as fibre orientation since these in turn strongly influences the quality of the moulded products in the injection moulding process. Numerical examples are provided for different flow parameters and the results obtained are analysed and discussed.

## 2. Governing Equations

The conservation of mass and momentum for an isothermal flow of fibre suspension can be expressed by the following equations,

$$\nabla \cdot \mathbf{u} = 0, \quad \rho\left(\frac{\partial \mathbf{u}}{\partial t} + \mathbf{u} \cdot \nabla \mathbf{u}\right) = -\nabla p + \nabla \cdot \boldsymbol{\tau}_f + \mu \nabla^2 \mathbf{u} + \rho \mathbf{g}, \tag{1}$$

where $\rho$ is the fluid density, $\mathbf{g}$ the gravity acceleration, $t$ the time, $\mathbf{u}$ the velocity vector, $p$ the hydrodynamic pressure, and $\boldsymbol{\tau}_f$ the stress tensor from the fibre suspension, and $\mu \nabla^2 \mathbf{u}$ the stress tensor from the Newtonian fluid with $\mu$ being the fluid viscosity.

The constitutive equation for the fibre stress is (Phan-Thien and Graham 1991; Fan et al. 1999),

$$\boldsymbol{\tau}_f = \mu\, f(\varphi)[A\dot{\boldsymbol{\gamma}} :<\mathbf{pppp}> + 2D_r F <\mathbf{pp}>], \tag{2}$$

where

$$f(\varphi) = \frac{\varphi(2-\varphi/\varphi_m)}{2(1-\varphi/\varphi_m)^2}, \quad A = \frac{a_r^2}{2(\ln 2a_r - 1.5)}, \quad F = \frac{3a_r^2}{\ln(2a_r - 0.5)}. \tag{3}$$

Here $\dot{\boldsymbol{\gamma}} = (\nabla \mathbf{u}^T + \nabla \mathbf{u})/2$ is the rate-of-strain tensor, $\mathbf{p}$ a unit vector along the axis of the fibre, $D_r$ the diffusion coefficient, $<...>$ denotes the ensemble average over the orientation space of fibres, $<\mathbf{pp}>$ and $<\mathbf{pppp}>$ are the second and forth-order structure tensors, respectively, and $\varphi$ the volume concentration. The parameter $\varphi_m$ express as the maximum volume packing of fibres, and it can be approximated by the following empirical linear function of the aspect ratio $a_r$,

$$\varphi_m = 0.53 - 0.013 a_r, \quad 5 < a_r < 30. \tag{4}$$

In Eq.(2), the first term on the right hand side expresses the stress contribution from the fibre motion, rotation and interactions (dissipative), and the second term represents the stress contribution due to the momentum transport caused by random motion of fibres. Generally, the first term plays a dominant role over the second term.

The second and fourth order structure tensors of fibres in Eq.(2) have to be calculated after the unit vector $\mathbf{p}$ is solved for each fibre. The evolution of the unit vector $\mathbf{p}$ is expressed by the Jeffery equation (Jeffery 1922),

$$\frac{d}{dt}\mathbf{p}(i) = \mathbf{L} \cdot \mathbf{p}(i) - \mathbf{L} : \mathbf{p}(i)\mathbf{p}(i)\mathbf{p}(i) + (\mathbf{I} - \mathbf{p}(i)\mathbf{p}(i)) \cdot \mathbf{F}^{(b)}(t). \tag{5}$$

Here, a diffusion term has been added to Jeffery equation according to Folgar-Tucker (Folger and Tucker 1984). In Eq.(5), $\mathbf{p}(i)$ is the unit vector along the axis of the $i$th fibre, $L$ the effective velocity tensor, $\mathbf{L} = \nabla \mathbf{u}^T - \varsigma \dot{\boldsymbol{\gamma}}$, with $\varsigma = (a_r + 1)^{-1}$. $\mathbf{F}^{(b)}(t)$ is a random force, with properties

$$\left\langle \mathbf{F}^{(b)}(t) \right\rangle = 0, \tag{6}$$

and

$$\left\langle \mathbf{F}^{(b)}(t+s)\mathbf{F}^{(b)}(t) \right\rangle = 2D_r \delta(s)\mathbf{I}, \tag{7}$$

where $\delta(s)$ is the Dirac delta function, $\mathbf{I}$ is the unit tensor, and $D_r$ is a diffusion factor also mentioned in Eq.(2). Folgar and Tucker assumed that

$$D_r = C_i \dot{\gamma}, \tag{8}$$

where $\dot{\gamma} = \sqrt{tr\dot{\gamma}^2/2}$ is the generalized strain rate, and $C_i$ is the interaction factor which can be a function of $\varphi$ and $a_r$. Phan-Thien et al. (2002) have provided a semi empirical equation for $C_i$. The random force can be expressed in terms of the white noise,

$$\mathbf{F}^{(b)}(t) = \sqrt{2C_i\dot{\gamma}} \frac{d\mathbf{w}_t}{dt}, \tag{9}$$

where $\mathbf{w}_t$ is the Wiener process and it is a Gaussian random function.

After **p**(i) is solved from Eq.(5), one can calculate the structure tensor using the ensemble average:

$$\left\langle \mathbf{pp} \right\rangle = \frac{1}{N}\sum_{i=1}^{N} \mathbf{p}(i)\mathbf{p}(i), \quad \left\langle \mathbf{pppp} \right\rangle = \frac{1}{N}\sum_{i=1}^{N} \mathbf{p}(i)\mathbf{p}(i)\mathbf{p}(i)\mathbf{p}(i), \tag{10}$$

where N is the number of fibres. Next, we introduce

$$\mathbf{q}(i) = q(i)\mathbf{p}(i) \tag{11}$$

with $q(i)$ being the modulus of $\mathbf{q}(i)$. Equation (5) then becomes

$$\frac{d}{dt}\mathbf{q}(i) = \mathbf{L} \cdot \mathbf{q}(i) + q(i)\mathbf{F}^{(b)}(t). \tag{12}$$

Rewriting Eq.(12), we have

$$\frac{\partial}{\partial t}\mathbf{q}(i) = -\mathbf{u}\cdot\nabla\mathbf{q}(i) + \mathbf{L}\cdot\mathbf{q}(i) + q(i)\sqrt{2C_i\dot{\gamma}}\frac{d\mathbf{w}}{dt}. \tag{13}$$

This equation can be solved for each fibre using a time marching scheme evolving with the flow field.

In order to properly describe the effect of the fibre suspension on the flow, a dimensional analysis is performed to yield various dimensionless group parameters. Regarding the right hand side of Eq.(2), we may obtain

$$<\mathbf{pppp}> \sim 1 \text{ and } \boldsymbol{\tau}_f \sim \mu\, f(\varphi) A \dot{\boldsymbol{\gamma}}. \tag{14}$$

Then, we define two Reynolds numbers to be given as

$$\text{Re}_1 = \frac{\rho U L}{\mu} \tag{15a}$$

and

$$\text{Re}_2 = \frac{\rho U L}{\mu f(\varphi) A}. \tag{15b}$$

The ratio of the two Reynolds numbers is

$$K = \frac{\text{Re}_1}{\text{Re}_2} = f(\varphi) A. \tag{16}$$

Strictly, the parameter $K$ represents the ratio of the viscosities due to the fibres and the solvent. It plays a rather similar role to the viscosity ratio of the solute to the solvent in a viscoelastic flow.

The second term on the right hand side of Eq.(2) represents the diffusion effect of fibre stress. The diffusion influence on the fibre stress can be expressed as

$$D = \frac{2D_r F}{A|\dot{\boldsymbol{\gamma}}|} = \frac{2C_i|\dot{\boldsymbol{\gamma}}|F}{A|\dot{\boldsymbol{\gamma}}|} = \frac{2C_i F}{A}. \tag{17}$$

For a given fibre aspect ratio, the parameter D in Eq.(17) is a constant. Therefore, D also reflects the influence of the aspect ratio of fibres on the stress diffusion. A higher value of D increases the gradients of stresses and in doing so, exerts a greater influence on the distribution of the velocity field.

The effect of surface tension on the flow behaviours can be expressed using the Weber number or the capillary number. For the Weber number,

$$We = \frac{U^2 \rho L}{\sigma}, \tag{18}$$

where $\sigma$ is the coefficient of surface tension (Weber number stands for the ratio of inertia force and surface tension force).

## 3. Level Set Method for Two-Phase Flows

The level set method is used to solve the following convective equation,

$$\frac{\partial \phi}{\partial t} + \mathbf{u} \cdot \nabla \phi = 0, \tag{19}$$

where $\phi$ is the level set function.

The interface between the two fluids is expressed by the null value of a function $\phi$; the function is positive in one phase and negative in another phase. This function is convected with the fluid flow via evaluating a hyperbolic equation as time progresses. Generally, this function is treated as a distance function from the interface. Using the formulation in the reference (Sethian and Smereka 2003; Sussman et al. 1994; Sussman et al. 1998), the level function is defined as,

$$\phi(\mathbf{x},t) = \begin{cases} > 0 & \mathbf{u} = \mathbf{u}_l & \text{if } \mathbf{x} \in \text{the liquid}, \\ = 0 & \mathbf{u} = \mathbf{u}_l & \text{if } \mathbf{x} \in \Gamma, \\ < 0 & \mathbf{u} = \mathbf{u}_g & \text{if } \mathbf{x} \in \text{the gas}, \end{cases} \tag{20}$$

where **u** is the "unified" fluid velocity. The abrupt change of fluid properties across the fluid-fluid interface such as density and viscosity may cause numerical difficulty in the solution of the governing equations. A finite thickness of the interface is defined to smooth the fluids properties (Sethian and Smereka 2003; Sussman et al. 1994; Sussman et al. 1998).

In the calculation, the level set function is initialized as a signed distance function from the interface. It is zero along the interface. Then, it is evolved according to Equation (19), which is solved by a time marching scheme. After each time step, the zero level set function should represent the new position of the interface. However, because of the numerical approximation, the level set function may not remain a distance function at later time steps; in particular after a long simulation time. Therefore, it is suggested that the level set function is re-initialized after every time step so that it remains a distance function without changing its zero level set. Generally, this is achieved by solving the following partial differential equation (Sethian and Smereka 2003; Sussman et al. 1994; Sussman et al. 1998),

$$\frac{\partial \phi}{\partial \tau} = sign(\phi)(1-|\nabla \phi|) \tag{21}$$

with initial conditions $\phi(x,0) = \phi_0(x)$, where *sign* is the sign function, $\tau$ is the (artificial) time, and $\phi_0(x)$ is the initial value of $\phi$ given at the beginning of calculation for all the domain. This equation is a nonlinear hyperbolic equation, and the solution of equation (21) reaching a steady state with the artificial time $\tau$ will be a distance function.

In order to increase the accuracy for keeping mass conservation during the re-initialization, Sussman et al. (1998) proposed the following improvement,

$$\frac{\partial \phi}{\partial \tau} = sign(\phi)(1-|\nabla \phi|) + \lambda_{ij} H'(\phi)|\nabla \phi| \tag{22}$$

where $\lambda_{ij}$ is a constant in each cell which is determined by keeping the volume conservation during the re-initialization.

The Navier-Stokes equations for two-fluid flows can be modified to include the surface tension force. Thus, the governing equation for the fluid velocity, **u**, along with the boundary conditions can be combined in a single equation as (Chang et al. 1996; Sussman et al. 1994)

$$\rho(\phi)\frac{D\mathbf{u}}{Dt} = -\nabla p + \nabla \cdot \boldsymbol{\tau}_f + \mu(\phi)\nabla^2 \mathbf{u} + \rho(\phi)\mathbf{g} - \sigma k(\phi)\delta(\phi)\nabla\phi,$$
$$\nabla \cdot \mathbf{u} = 0, \tag{23}$$

where $\delta(\phi)$ is the Dirac delta function, $\sigma$ is the surface tension factor, and $k(\phi)$ is the curvature of the interface. $\rho(\phi)$ and $\mu(\phi)$ are the density and viscosity, respectively, which can take on two different values depending on $\phi$, and can be expressed as

$$\rho(\phi) = \rho_g + (\rho_l - \rho_g)H(\phi) = \rho_l[c + (1-c)H(\phi)], \tag{24}$$

and

$$\mu(\phi) = \mu_g + (\mu_l - \mu_g)H(\phi) = \mu_l[b + (1-b)H(\phi)], \tag{25}$$

where $c = \rho_g / \rho_l$ is the density ratio and $b = \mu_g / \mu_l$ is the viscosity ratio. $H(\phi)$ is the Heaviside function given by

$$H(\phi) = \begin{cases} 0 & \text{if } \phi < 0 \\ \frac{1}{2} & \text{if } \phi = 0 \\ 1 & \text{if } \phi > 0 \end{cases}. \tag{26}$$

In order to reduce the numerical difficulties due to abrupt change of density and viscosity at the interface expressed by Eq.(24) to Eq.(26) and due to Dirac delta function in the surface tension term, the interface is smoothed and a finite thickness of interface is given which is proportional to the spatial mesh size (Chang et al. 1996; Sussman et al. 1994). The density and viscosity are expressed as

$$\rho_\varepsilon(\phi) = \rho_l[c + (1-c)H_\varepsilon(\phi)] \tag{27}$$

and

$$\mu_\varepsilon(\phi) = \mu_l[b + (1-b)H_\varepsilon(\phi)] \tag{28}$$

with

$$H_\varepsilon(\phi) = \begin{cases} 0 & \text{if } \phi < -\varepsilon \\ \frac{1}{2}\left[1 + \frac{\phi}{\varepsilon} - \frac{1}{\pi}\sin(\pi\phi/\varepsilon)\right] & \text{if } |\phi| \le \varepsilon \\ 1 & \text{if } \phi > \varepsilon \end{cases}. \tag{29}$$

The equivalent smoothed delta function is $\delta_\varepsilon(\phi) = \dfrac{dH_\varepsilon}{d\phi}$. The curvature of the interface can be calculated from the function $\phi(\mathbf{x},t)$,

$$k(\phi) = -\frac{\phi_y^2 \phi_{xx} - 2\phi_x \phi_y \phi_{xy} + \phi_x^2 \phi_{yy}}{\left(\phi_x^2 + \phi_y^2\right)^{3/2}}. \tag{30}$$

The Eq. (23) can be rewritten in the dimensionless form by normalizing all the parameters (Chang et al. 1996; Sussman et al. 1994),

$$\mathbf{u}_t + \frac{\nabla p}{\rho(\phi)} = \mathbf{F} \tag{31}$$

where

$$\mathbf{F} = -\mathbf{u} \cdot \nabla \mathbf{u} + \frac{1}{\rho(\phi)} \left( \frac{1}{\text{Re}_1} \nabla \cdot \boldsymbol{\tau}_f + \frac{1}{\text{Re}_1} \nabla \cdot (2\mu(\phi)\dot{\boldsymbol{\gamma}}) - \frac{1}{We} k(\phi)\delta(\phi)\nabla\phi \right), \tag{32}$$

and $\mathbf{e}$ is the unit vector of the gravitational force.

Bell and Marcus (1992) and Sussman et al. (1994) described a variable density projection method. In this method, it is assumed that

$$\nabla \cdot \mathbf{u}_t = 0. \tag{33}$$

Thus, according to the Hodge decomposition, one can uniquely decompose the quantity $\mathbf{F}$ in Eq.(31) into a divergence free part ($\mathbf{u}_t$) and the gradient of a scalar divided by density ($\nabla p / \rho(\phi)$). Since $\mathbf{u}_t$ is divergence free we can write it as for two-dimensional flow as

$$\mathbf{u}_t = (\partial \psi_t / \partial y, -\partial \psi_t / \partial x, 0)^T \tag{34}$$

where $\psi_t$ is the stream function corresponding to $\mathbf{u}_t$.

For two-dimensional flows, if we multiply both sides of Equation (31) by $\rho$ and take the curl of both sides, we then obtain,

$$-\nabla \cdot (\rho \nabla \psi_t)\mathbf{k} = \nabla \times (\rho \mathbf{F}), \tag{35}$$

where **k** is the unit vector in the *z* direction.

For given smooth boundary conditions and initial conditions, Eq.(35) can be solved for the stream function $\psi_t$ for a prescribed time increment. After this, the velocity time derivative can be obtained by Eq. (34). Then, the new velocity field is calculated using $\mathbf{u}^{n+1} = \mathbf{u}^n + \mathbf{u}_t \Delta t$.

## 4. Numerical Implementation: Discretization and Algorithm

The discretization is based on a staggered grid arrangement, as shown in Fig. 1. Here, $\mathbf{u}$, $\mathbf{q}$, $\rho$, $\mu$, $\phi$, are given at the primary grid points denoted by open circles, and $\psi_t$, $\nabla \mathbf{u}$, and $\boldsymbol{\tau}$ are defined on the dual grid points denoted by "×". Actually, the dual grid points lie on the wall boundary of the physical domain for the imposition of the no-slip boundary conditions (Fig. 1).

The convective terms in Eqs.(13), (19) and (32) are discretised using up-winding scheme with a high-order essentially non-oscillatory (ENO) procedure for the dependent variables u, v, **q** and $\phi$ at edges of each cell, $(i \pm 1/2, j)$ and $(i, j \pm 1/2)$, see Bell and Marcus (1992), Chang et al. (1996), Sussman et al. (1994), and Dou et al. (2004). The formulation for the convective term for the variable $\phi$ is as given in Bell and Marcus (1992), Sussman et al. (1994), Sussman et al. (1998), and Dou et al. (2004):

$$\mathbf{u} \cdot \nabla \phi = u_{ij}\left(\phi_{i+1/2,j} - \phi_{i-1/2,j}\right)/\Delta x + v_{ij}\left(\phi_{i,j+1/2} - \phi_{i,j-1/2}\right)/\Delta y . \qquad (36)$$

The discretization of u, v, and **q** used a similar form as that for $\phi$. The viscous term and the fibre stress term in Eq. (32) are discretized using the central differencing method (Bell and Marcus 1992; Sussman et al. 1998).

For the discretization of the surface tension $-\frac{1}{We}k(\phi)\delta(\phi)\nabla\phi$ in Eq.(32), this term is simplified via $\nabla H(\phi) = \delta(\phi)\nabla\phi$, where $H(\phi)$ is the Heaviside function as defined in Eq. (29). The resulting contribution to the right hand side of Eq. (35) due to surface tension is thus

$$-\frac{1}{We}\left(k_x H_y - k_y H_x\right) \qquad (37)$$

The derivatives of *k* and *H* are discretized using central differencing.

The projection Eq.(35) is rewritten below using the difference operators for the divergence and gradient,

$$-D_x(\rho G_x \psi_t) - D_y(\rho G_y \psi_t) = D_x(\rho F_y) - D_y(\rho F_x). \tag{38}$$

Here, the discrete form of the divergence ($D$) and gradient ($G$) operators are (Bell and Marcus 1992; Sussman et al. 1998):

$$D_x f = (f_{i+1,j+1} - f_{i,j+1} + f_{i+1,j} - f_{i,j})/(2\Delta x) \tag{39}$$

$$D_y f = (f_{i+1,j+1} - f_{i+1,j} + f_{i,j+1} - f_{i,j})/(2\Delta y) \tag{40}$$

$$G_x f = (f_{i+1/2,j+1/2} - f_{i-1/2,j+1/2} + f_{i+1/2,j-1/2} - f_{i-1/2,j-1/2})/(2\Delta x) \tag{41}$$

$$G_y f = (f_{i+1/2,j+1/2} - f_{i+1/2,j-1/2} + f_{i-1/2,j+1/2} - f_{i-1/2,j-1/2})/(2\Delta y). \tag{42}$$

The projection Eq. (38) is solved for the discrete scalar $\psi_t$, which is defined at cell corners (i+1/2,j+1/2), using a preconditioned conjugate gradient (PCG) method. Then, $\mathbf{u}_t$ is obtained via Eq.(34) using the central difference scheme. The time marching of the dependent variable u, v, and $\phi$ are calculated via a high order Total Variation Diminishing (TVD) Runge-Kutta scheme. The time step $\Delta t$ is determined by restrictions due to CFL condition, viscosity and surface tension (Sussman et al. 1998; Dou et al. 2004).

In addition, in order to increase the accuracy in the calculation of the velocity gradient at the wall, a special extrapolation technique is used. This additional extrapolation ensures the solution at the wall to be second-order accurate. This correction also improves the accuracy of the calculation of stress gradient and enhances the numerical convergence and accuracy (Dou et al. 2004).

Overall, the total numerical procedure can be summarized as below:
(1) Initialize all the parameters. Give the initial $\phi$ in the domain.
(2) For given $\phi^n$, $\mathbf{u}^n$, $\rho(\phi)$, $\mu(\phi)$ at the nth time step, solve for Eq.(13) using the time marching scheme, and obtain the distribution of fibre orientation ($q_1, q_2, q_3$) and the configuration tensor $<\mathbf{pp}>$ and $<\mathbf{pppp}>$.
(3) Calculate $\rho(\phi)$ and $\mu(\phi)$ from Eqs.(27) and (28), respectively. Next, evaluate the convective term (ENO scheme), viscous term and fibre stress term (central difference), source term (coordinate components), and tension term in Eq.(32). Then, evaluate the value of **F** by summing the above terms. Simultaneously, evaluate the convective term (ENO scheme) in Eq.(19) for later use.

(4) Solve for Eq.(35) using a preconditioned conjugate gradient (PCG) method to obtain $\psi_t$. Calculate $\mathbf{u}_t$ from Eq.(34) by differencing the $\psi_t$.

(5) Determine the time step $\Delta t$ using the criteria given. Do the time marching step for variables u, v, and $\phi$ using a high order total variation diminishing (TVD) Runge-Kutta scheme (3$^{rd}$ order) to obtain at the next (n+1) time level $u^{n+1}$, $v^{n+1}$, and $\phi^{n+1}$.

(6) Re-initialize the level set function for $\phi^{n+1}$ via Eq. (22).

(7) Return to step (2) and evaluate until a prescribed time.

## 5. Simulation Results and Discussions
### 5.1 Conditions of simulations

A rather annoying feature in the application of the projection/level set method is that it is required that the initial condition for the computational domain be divergence free. This requirement is only satisfied for few situations. Unfortunately, for most engineering problems, this condition is not satisfied, such as that encountered in this study. In the present work, a simplified initial condition is employed. First, we assign a continuous smooth distribution of stream function. Then, a divergence free velocity field is obtained by differentiating this stream function and is taken as the initial flow field. For convenience, a uniform plug flow or a parabolic distribution of velocity can be assumed for the flow in a duct.

The computing domain is shown in Fig.2 as a rectangular area with inlet AD, and BC is the outlet. AB and CD are solid walls and EF is the interface of the two fluids. AB is taken as 4 times of AD, and AF is taken as 0.25 times of AD. The liquid fluid flow is driven by a pressure gradient at the inlet in a two-dimensional channel. Initially, the liquid fluid stays in the area enclosed by AFED and the gas fluid (air) fills the area enclosed by EFBC. The shape of the flow front at t=0 is a straight line EF. At time *t=0*, the liquid fluid is allowed to flow as a given velocity distribution and the flow front becomes a curved line convex to the negative y direction. At the first few steps, the flow will automatically adjust itself to obey the governing equations. After a sufficiently long enough time, the effect of initial flow field becomes small. Along the channel the flow front evolves even as the flow is influenced by the pressure gradient, fluid convection, and viscous and fibre forces as well as the surface tension.

For the flow of a Newtonian fluid, it can be assumed that the velocity distribution in the duct far from the front obeys a parabolic distribution, in which the effect of interface and surface tension can be neglected. In the simulation, it is our intent to track the evolution of the flow front, and to study the effect of the fibre suspension properties. The boundary condition is taken such that AB and CD are solid surfaces and hence no-slip boundary condition applies. Boundary conditions at inlet for velocities are given at AD and Neumann boundary condition applies on BC assuming the duct is reasonably long. The boundary condition on inlet AD is given as parabolic. When the flow front is far from the inlet, the effect of inlet condition on the front shape and fountain flow is small. At the wall, the level set function is extrapolated to the wall from the interior domain. The slope of the interface at the wall is directly obtained from the simulation. The boundary condition of **p** at the wall is given as parallel to the wall. The initial condition for **p** (fibre orientation) within the domain is given to be random in the whole domain (ABCD in Fig.2). Although there should be no fibre in the air area, fibres are assigned in the calculation at all the nodes for the purpose of the stress evaluation as the level set method required. With the fluid flowing and as the time progresses, the orientation of the fibres will automatically adapt to the velocity field.

The simulation is carried out for given concentration and number of fibres. These are set to be constants in the domain as in the work of Hulsen et al. (1997) and Fan et al. (1999). In another word, the concentration of fibres is set to be uniformly distributed in the whole flow domain (i.e., no migration exists in the domain). Fan et al (1999) also employed such a uniform distribution of fibre concentration for their simulation. In a further two additional cases, this constraint is removed and simulation of non-uniform concentration distribution is studied. Generally, there is no limitation on the number of fibres in principle, provided that the computer resources are sufficiently large to allow for such calculation. Equation (5) is solved for each fibre, and then ensemble average from all the fibres is employed to calculate the stresses at each node. It is suggested that a sufficiently large number of fibres are employed in the simulation to avoid oscillations in the calculations; on the other hand, beyond a certain large number of fibres utilized, the simulation results remain fairly consistent.

Overall, where there is strong shear and surface tension, or local concentration of fibres, the associated stress gradient remains high; these characteristics can be obtained near the wall and the interface (Fan et al. 1999). Numerical limitation is encountered when the concentration is very high or the aspect ratio is very large which leads to very high stress gradients near the interface and the walls. These large stress gradients greatly impede solution of the momentum equations to a higher concentration when the other parameters are given. We simulated for a number of runs as the fibre parameter variation. Listed in the Table 1 are those typical runs made and to be used for further discussion in the following sections.

| Run No. | φ | $a_r$ | $Re_1$ | $K=Re_1/Re_2$ | D | We |
|---|---|---|---|---|---|---|
| 1 | 0.05 | 10 | 300 | 2.05 | 0.072 | 3.044 |
| 2 | 0.10 | 10 | 300 | 5.20 | 0.072 | 3.044 |
| 3 | 0.15 | 10 | 300 | 10.43 | 0.072 | 3.044 |
| 4 | 0.25 | 10 | 300 | 40.86 | 0.072 | 3.044 |
| 5 | 0.0 | 0 | 300 | 0 | | 3.044 |
| 6 | 0.05 | 20 | 300 | 6.24 | 0.082 | 3.044 |
| 7 | 0.10 | 20 | 300 | 18.78 | 0.082 | 3.044 |
| 8 | 0.15 | 20 | 300 | 50.11 | 0.082 | 3.044 |
| 9 | 0.10 | 10 | 300 | 5.20 | 0.072 | 5.555 |
| 10 | 0.10 | 10 | 300 | 5.20 | 0.072 | 11.11 |
| 11 | 0.10 | 10 | 300 | 5.20 | 0.072 | 3.044 |
| 12 | 0.25 | 10 | 300 | 40.86 | 0.072 | 3.044 |
| 13 | 0.10 | 10 | 167 | 5.20 | 0.072 | 3.044 |
| 14 | 0.10 | 10 | 67 | 5.20 | 0.072 | 3.044 |
| 15 | 0.10 | 10 | 0 | 5.20 | 0.072 | 3.044 |

Table 1 Flow parameters used for the simulations. The number of fibres at each node is $N=1000$ and the interaction coefficient is $C_i = 0.01$. Run No.11 and Run No.12 give results with non-uniform distribution of fibre concentration (see Eq.(43)).

## 5.2 Flow Front evolution

The simulated results of flow front moving with the time evolution are shown in Fig.3 for two typical examples which show the influence of fibre concentration. The time in the figure is expressed with dimensionless. The time t=1 means t=100 $\Delta t$ and t=5 means t=5×100 $\Delta t$ and so on; the time step $\Delta t$ is that used for the time marching according to the Runge-Kutta scheme. Therefore, the time interval between two contour lines is 200 $\Delta t$. Figure 4 shows the comparison of flow front for different cases at time t=27. In these figures, Reynolds number $Re_1$ based on the averaged velocity at inlet, the channel width, and the viscosity of the solvent is provided as in Table 1. The grid employed is 27×102. Prior to that, grid invariance tests were carried out with 17×62, 27×102 and 37×142. For the grid of 37×142, the number of fibres is subjected to the limit of the memory of the computer, but it is good for small number of fibres. The number of fibres used in this study is 1000 which is enough large for the material parameters employed. The dimensionless time interval for contour of the flow front in Fig.3 is 2.0. In all the cases listed in Table 1, the volumetric flow rate at the inlet is prescribed, which is defined as the integration of the velocity along the channel width. The fluid flow is from top to bottom in Fig.3. For the geometry shown in Fig.2, a program running for a given set of fibre parameters typically takes about 3000 time steps. In the initial stage of flow development, the flow front varies largely with time partly due to the unsteady fibre stresses. After the flow front moves for a sufficiently long distance, the shape of the flow front becomes nearly constant in shape. For example, for a Newtonian flow shown in Fig.4 (Run No.5), the initial transient stage is up to at t=7, while for the fibre suspension flow with $\varphi = 0.25$ and $a_r = 10$ as depicted in Fig.3 and Fig.4 (Run No.4) this initial stage is limited to t=9. The factors influencing the flow front are the viscous stresses, stresses due to fibre orientation, pressure gradient, convective inertia, and surface tension. Figure 3 show the influence of fibre concentration on the development of flow front for aspect ratio $a_r = 10$. It is observed that the effect of fibre concentration on the front shape is not significant. Effect of surface tension on the front shape can be observed from Fig.4, and it seems that surface tension has a relatively larger influence on the front shape. When the surface tension is large, the flow front seems much fuller (Fig.4, Run No.2), while a low surface tension leads to a little longer front in the streamwise direction (Fig.4, Run No.9). From Fig.4, it is found that the aspect ratio has little influence on the front shape for aspect ratio varying from $a_r = 10$ to $a_r = 20$. It is also found from Fig.4 that there is no essentially difference for the front shape for the pure Newtonian fluid flow without fibres (Run No.5) and the flow with fibre suspension at same surface tension (Run Nos. 2 and 4). From Fig.4, it is observed that the flow front as a whole is near a semi-circle for Newtonian case. There is some difference near the boundaries with the variation of fibre parameters. In the fibre suspension flow, the variation of the front shape is attributed to the fibre motion. The contributing influence of the fibre on the velocity profile is through the non-Newtonian stresses produced by the fibre motion. These said stresses are concentrated in the narrow band near the boundary. Therefore, the influence of fibre parameter as depicted in these figures is found in the region near the boundary. In the central core region, the fibre stresses exhibit no big change along the width and hence the effect of fibre on the front shape is small. In conclusion, the influence of variation of fibre parameter on the front shape is limited at the fixed Re.

In summary, as shown in Fig.3 and Fig.4, effect of the variation of fibre parameter on the front shape is very small at Re=300. The influence of fibre stress is mainly focused or concentrated in the region near the wall. An increase of the fibre concentration makes the interface front more tangent to the wall for the region near the wall. Even though this variation of fibre stress can be large in the streamwise direction, it is not easily/explicitly reflected by the front shape; this is because the change in the transverse direction is relatively small. In the central core area, the effect of fibre stress is small. Hence, the variation of fibre parameter has little effect on the front shape.

Figure 5 shows a picture during the molding of sandwich materials in an experiment at different times (Nguyen-Chung and Menig 2001). In this experiment, two different materials were used to track the interface to see how the fountain flow develops with time. It is broadly observed that the front shape photographed in the experiments agrees reasonably with the present simulations at least qualitatively.

**5.3 Velocity vector and streamlines near the front**

For the flow with front moving in the duct, a "fountain flow" generally formed near the front due to the no-slip condition at the walls. The "fountain flow" means that the fluid near the front spread out as the flow moves forward (like a spring extending). The distributions of velocity vectors for two sets of fibre parameters are shown in Fig.6 as typical examples. The fountain flow near the flow front is unmistakeable. There is a transverse velocity on the both sides of the centreline near the front area. In this study, it is small compared with the main flow. It is less than 10% of the streamwise velocity and the difference of the transverse velocity in fibre suspension flow with that in the Newtonian flow is less than 6%. Since the magnitude of the transversal velocity is so small compared to the streamwise velocity, it is difficult to distinguish how the fibre parameter influences the variation of velocity vector.

On the other hand, it is observed that the front slope near the wall for Run No.4 is steeper than the other case, which is due to the high stresses for the high concentration with fibres with high aspect ratio. This phenomenon is a hallmark of the fibre orientation influence at high concentration. When the concentration is low, the front shape with fibre suspension has no big difference from the (usual) Newtonian flow. The exact magnitude of fountain flow can be found in the velocity distribution along the width. The detailed velocity distribution under the influence of fibre motion will be discussed in later sections.

To artificially make the transversal component of the velocity comparable with the axial component of the velocity near the front, the streamlines in a moving frame which is fixed on the fore front are plotted in Fig.7 as typical examples. It can be now seen how the fountain flow varies near the front. Comparing Fig.7(a) with Fig.7(b), it is found that the fountain flow is reduced with the increasing fibre concentration. It is also found that at far upstream of the flow front, the position of zero streamwise velocity in the moving frame migrates towards the wall with the increase of the fibre concentration. This means that the streamwise velocity is increased near the wall for high fibre concentration.

In summary, the influence of fibre addition to the flow on the velocity vector field in the fountain flow is small as seen in the natural coordinates. This is because the cross flow is generally small in the fountain flow area. However, in a moving frame which is fixed at the flow front, the influence of fibre parameter on the fountain flow is clearly seen. These variations of velocities in the fountain flow may have certain effects on the fibre orientation and the quality of the resulting products.

**5.4 Fibre orientation in the mold filling**

In this section we present the simulation results for the orientation of fibres in the flow field. In the simulation with a model in which the fibre is aligned with the velocity vector, the fibre orientation is always along the streamlines. On the other hand, the orientation obtained by solving the Jeffery equation like in this study can be represented by an ellipsoid whose major axis is in the direction of the orientation. The eigenvectors and eigenvalues of the structure tensor **A** indicate the orientation directions and the relative magnitude of the alignment with respect to these directions. Figures 8-9 show the fibre orientation with the time evolution for various fibre parameters. In these pictures, a line is drawn at each node in the direction of the eigenvector associated with the largest eigenvalue of **A**, with the length of the line proportional to the eigenvalue. As a result, the direction of each line indicates the main direction of local orientation, and the length of the line shows the extent of alignment with the associated direction.

It should be mentioned that although the flow is two-dimensional, the fibre motion is three-dimensional. Thus, the structure tensor **A** is third-order. In Figs 8-9, it can be seen that the major axis of the ellipsoid of the tensor **A** lies on the x-y plane which is associated with the first maximum eigenvalue. In these figures, in the area enveloped with thin lines at the front adjacent to the wall, the plotted is the direction of the eigenvector with the associated second maximum eigenvalue. In these areas, the direction of the eigenvector with the associated first maximum eigenvalue is along the z axis which is perpendicular to the x-y plane. An explanation for this phenomenon is as follow. In the area at the corner of the front adjacent to the wall, the flow is subjected to a large deformation, which can be observed from Fig.6 and Fig.7. This deformation leads to a large vorticity which is along the z direction. Thus, the fibre orientation is adapted to the vorticity field and therefore the major axis of the tensor **A** becomes alignment with the vorticity axis. This analysis is in agreement with those in Harlen and Koch (1997).

These distributions of fibre orientation in Figs.8-9 are calculated using ensemble average for the total number of configuration field at each node and for during a time interval which consists of about 100 time steps (100 $\Delta t$) in order to view the fibre orientation developing along the streamwise channel length (the time step $\Delta t$ is that used for the time marching according to the Runge-Kutta scheme). These averaged fibre orientations are strictly still unsteady (instantaneous), and evolve with time. In these figures, the liquid is above the flow front while the air is below the flow front. The fibres in the air area have no physical meaning for the displayed results; these are necessary as an artifice for the calculation in the numerical scheme as explained earlier in previous sections. The level set method requires the same numerical computation be carried out for both the liquid and the air area because it utilizes the unified formulations and solutions.

Thus, in the air area, computation for fibre is also carried out through even though there is no fibre in the air area. The effect of considering fiber in the air region on the simulation results is very small and this can be neglected. This is because the ratio of the densities between air and liquid is about 1/1000. Thus, this treatment only leads to an error of 0.001 on the shape of the front calculation. This error estimate can be easily deduced form Eq. (27). In all these figures, the fibre could be removed when the pictures are plotted. However, they are kept in the figures in order to reflect the real state of fibre at the interface after the simulation.

In the calculations, the concentration is taken as uniform in the whole field (except for Run.No.11 and 12). The number of the configuration number is constant at all the grid points. At each grid point, the orientation tensor **A=<pp>** is obtained by calculating ( $a_{11} = \frac{1}{N}\sum_{i=1}^{N} p_1 p_1$ , $a_{22} = \frac{1}{N}\sum_{i=1}^{N} p_2 p_2$ , $a_{33} = \frac{1}{N}\sum_{i=1}^{N} p_3 p_3$ , $a_{21} = a_{21} = \frac{1}{N}\sum_{i=1}^{N} p_1 p_2$ ) after solving Eq.(13) for the spatial orientation of each fibre. It can be observed from Figs.8-9 that the fibre orientation is basically symmetrical with the centreline of the channel. It can also be observed that the fibre orientation is not always aligned with the velocity vector, compared with Figs.6 and 7. This is manifested by the fibre being nearly aligned with the velocity vector at the upstream of the flow front and near the walls, but the fibre is always nearly "criss-crossing" or traversing the velocity vector at large angle along the centerline and in the region of fountain flow. From the Jeffery equation, the motion of a fibre is composed of convection, rotation and diffusion and interaction.

In the configuration field, the variation of the orientation of each fibre with the time is very large due to said effects of convection, rotation and diffusion. In particular, when the concentration is high and the aspect ratio is large, the variation of fibre orientation is faster with time. Thus, the structure tensor at each node is varying with time, and the stresses at each node are strictly oscillatory. However, when the number of configuration field is large enough, the ensemble average of the configuration field makes the oscillation relatively far smaller. Therefore, the value of the stresses in the whole domain can be considered as in a quasi-steady state.

The fibre orientation simulated and presented in Figs.8-9 can be explained as follows. The motion of a fibre is controlled by three forces namely, convection in the downstream direction due to the streamwise velocity, rotation due to shear rate, and diffusion due to the Wiener process owing to the non-uniformity of the fibre orientation and the non-uniformity of the strain rate distribution. For each time step, each component ($q_1$, $q_2$, and $q_3$) of fibre orientation **q** varies and it entails a variation due to the different roles of convection, rotating, and diffusion. Thus, the stress tensor resulting from the motion of fibre (**q**) varies with each time step, and the magnitudes of the stresses vary with the time step taken, i.e., $\tau_{xx}$, $\tau_{xy}$, and $\tau_{yy}$ are all unsteady activities. This contributes to the unsteady velocity **u** as evaluated from the momentum equations. Furthermore, the strain rate ($\dot{\gamma}$) is unsteady due to the unsteady **u**. In turn, the unsteady behaviour of the velocity gradient enhances the unsteadiness of the fibre convection, rotating and diffusion. The overall result is the increase of the unsteadiness of the fibre motion (**q**). When the dimensionless parameter K increases, the fibre stresses increase corresponding. As such, the unsteadiness of **q** resulting from the increase of the magnitude of fibre stresses becomes very large. However, when the number of configuration filed is very large, although the motion of single fibre is unsteady, the ensemble average of the average behaviour tends to steady (see next section). The above is a detailed description on how the fibre motion interacts with the fluid flow field.

**5.5 Effect of number of configuration field**

Figures 10 shows the effect of the number of Brownian configuration fields on the accuracy of simulations for Run No.4. The values of the normal stress $\tau_{f_{xx}}$, the normal stress $\tau_{f_{yy}}$, the shear stress $\tau_{f_{xy}}$ and the first normal stress difference $N_1 = \tau_{f_{yy}} - \tau_{f_{xx}}$ are recorded at the fixed location G in Fig.2 (the coordinates of position G is x/h=0.92, y/h=-0.24, and h is the half width of the channel) and are shown in Figs. 10(a), (b), (c), and (d), respectively. Since the streamwise direction is along the y axis and the transverse direction is along the x axis, the first normal stress difference is expressed as $N1 = \tau_{f_{yy}} - \tau_{f_{xx}}$. The numbers of Brownian configuration fields employed in these pictures are N=50, 500, 1000, and 2000, respectively. All of these figures clearly depict that there is a starting period before the flows assume the quasi-steady state. For the mesh employed, the minimum time for this period is about t=10. After this period, the stresses tend to almost a constant, although there still is some oscillation. It is observed from Fig.10 that the values of the stresses converge to constant values with the increase of N.

These observations suggest that increasing the number of Brownian configuration fields enhances the accuracy of the computation, although it may be mentioned that there is no marked or significant difference between N=1000 and N=2000. As such, it can be accepted that most of the calculations carried out in this work are for N=1000 without loss of fidelity. Finally, for the first normal stress difference $N_1$ which is the difference between Fig.10(a) and (b) and depicted in Fig.10(d), it may be observed that there is relatively large difference for N=50 and N=500 while N=1000 and N=2000 show much smaller variation and almost indistinguishable from each other. Therefore, all the simulated results in this study using N=1000 is acceptable.

**5.6 Distributions of stresses along the channel width**

Figures 11 shows the distributions of the stresses along the channel width at the position just behind the front (I-I line in Fig.2) for various concentrations for the aspect ratio $a_r = 10$. The normal stress $\tau_{f_{xx}}$, the normal stress $\tau_{f_{yy}}$, the shear stress $\tau_{f_{xy}}$, and the first normal stress difference $N_1 = \tau_{f_{yy}} - \tau_{f_{xx}}$ are depicted for concentration $\varphi$=0.05, 0.10, 0.15 and 0.25, respectively. From Fig.11(a) and (b), it is observed that both the normal stress in transverse direction and the normal stress in streamwise direction increase with the increasing concentration, while the former has a maximum at the centreline and the latter obtains its maximum at the wall. These phenomena are due to the fact that the fibre orientation is near to be parallel to the wall when it approaches the wall and the fibre is near to cross the flow direction when it locates at the centreline. From Fig.11(c), it can be seen that the shear stress increases when it leaves the centreline, and but finally decreases to zero at the wall. This is because that the fibre orientation must be parallel to the wall at the wall.

Finally, it is observed from Fig.11(d) that the first normal stress difference obtains its maximum at centreline and reaches its minimum at the wall at the centreline. At high concentration, the first normal stress difference has a negative value at the centreline. This means that the fibre at the centreline in the region of fountain flow is more extended along the transverse direction at high concentration. It seems that there is large gradients of the stresses in the flow near the wall which are the sources to set the limit of numerical solution. On the numerical aspect, at high concentration, the steep stresses near the wall are hardly resolved by ordinal numerical methods. In order to resolve numerical stability at high stresses, Fan et al. employed an adaptive viscosity method to achieve the solution at high concentration, but with some sacrifice of accuracy at the region of high stresses (Fan et al. 1999).

At a location where it is far from the flow front (e.g., at the location G in Fig.2), the outline of the distributions of the normal stress $\tau_{f_{yy}}$ and the shear stress $\tau_{f_{xy}}$ are basically the same as those shown in Fig.11, respectively. However, the shape of the distribution of the normal stress $\tau_{f_{xx}}$ (not shown here) is different from that in Fig.11 (a). The maximum of the normal stress $\tau_{f_{xx}}$ is not at the centreline, but it locates at the both sides of the centreline symmetrically and has a profile like "M" at high concentration. In addition, the magnitude of the normal stress $\tau_{f_{xx}}$ is much smaller than that in Fig.11 (a). These differences of the phenomena can be obviously attributed to the influence of the fountain flow. In the region of fountain flow, the fibre is more oriented cross the main flow direction than that in the region far behind the front. As such, the fibre motion within the fountain flow with the increase of concentration eventually leads to the normal stress $\tau_{f_{xx}}$ to get its maximum at the centreline. All the phenomena discussed above accord with the physics of the fibre suspension described by the Jeffery equation and the rheology of fibre suspension.

**5.7 Velocity distributions near the front**

Figures 12-16 show the velocity distribution along the channel width just behind the flow front; as depicted figuratively in Fig.2 at the first grid line next to the intersection point of the flow front with the wall and marked as I-I cross the channel. These mentioned figures are for the cases of aspect ratio $a_r = 10$ and $a_r = 20$. The flow parameters are taken at the time t=25. From Figs.12(a-b), it is observed that the increase of fibre concentration reduces the streamwise velocity around the centreline and increases the streamwise velocity near the wall, and the increase of concentration reduces the transverse velocity. From these figures, it is clear that the transverse velocity for the pure Newtonian fluid flow is the largest which is about 10% of the streamwise velocity. For fluid parameters simulated in Fig.12, the maximum of the difference in transverse velocity from Newtonian case is about 6%. These changes of velocity distribution are due to the influence of distribution of fibre stresses. In the simulation, it is found that the distribution the first normal stress difference increases toward the wall across the width of channel. Therefore, it is suggested that this kind of stress distribution tries to "flatten" the streamwise velocity, and results in a reduction of the transverse velocity magnitude. Near the walls, the velocity profile becomes steeper with the increase of the concentration of fibres. Hence, the addition of fibres increases velocity gradient and the drag force in the channel flow and results in a large pressure drop. As the result, the pressure drop in the duct is increased with the increase of fibre concentration. These influences of velocity distribution and drag force increase are similar to that commonly and typical observed events regarding turbulence intensity in Newtonian flow.

From Fig.13, it is observed that the increase of aspect ratio reduces the streamwise velocity around the centreline and increases the streamwise velocity near the wall; the increase of aspect ratio reduces the transverse velocity. Thus, the effect of increasing the aspect ratio is the same as increasing the concentration. This is because the results of increasing the concentration or the aspect ratio both lead to the large increase of the fibre stresses. This is also rather obvious from Eq.(2) and Eq.(3).

Figure 14 shows the effect of surface tension on the velocity distributions. It is found that surface tension has little influence on the streamwise velocity behind the front. It is also seen that the decrease of surface tension reduces somewhat the transverse velocity. Thus, the decrease in surface tension decreases the strength of the "fountain." In the fountain flow area, the decrease of tension force is equivalent to the increase of contribution from the fibre stress. This will lead to the reduction of transversal flow (Fig.14). The decrease of Re makes larger the contribution from the fibre stresses. This will also lead to the reduction of transversal velocity (see subsequent figures).

In all the above Figs.12-14, the concentration of fibres is set to be uniform distribution in the whole flow domain (i.e., no migration is allowed or exists in the domain). The experimental data for the concentration distribution in the duct is very scanty in the literature for one to make a prescription for comparison and further study. The only work, however, by Yasuda et al. (2004) showed that the concentration is actually not quite truly uniform in both the dilute suspension and concentrated suspensions using glass fibres in a channel flow. The aspect ratio of the fibres employed by Yasuda et al. is about 60 and the concentration of fibres in the PB fluid is about 4% in volumetric fraction. Their measurement showed that the concentration is low near the walls and it is practically zero at the walls. In the center region away from the walls (x/h<0.8), the distribution of concentration is about uniform. This measurement is in somewhat agreement with the flow physics because there is indeed no fibre on the walls. Yasuda et al. commented that the weight center of fibres near the wall is away from the wall. The experiment indicated that there is a flow induced fibre orientation, and concentration distribution is not uniform in the fully developed channel flows. In order to see whether the non-uniformity of fibre concentration has an influence on the flow behaviour and flow front shape, further calculation is carried out for a prescribed distribution of concentration along the channel width,

$$\left.\begin{array}{ll} \varphi = \varphi_0 & for \quad x/h < 0.8 \\ \varphi = \varphi_0(1 - x^2/h^2) & for \quad x/h \geq 0.8 \end{array}\right\} \quad (43)$$

where $\varphi_0$ is the constant concentration distribution. In the calculation, the value of $\varphi_0$ at each node is kept constant. This result of velocity distributions are shown in Fig.15 (Run Nos.11 and 12). It is observed that the streamwise length of the flow front becomes slightly shorter when this non-uniform distribution is used on comparing to the counterpart in Fig.3 (Run Nos.2 and 4), i.e., the front near the wall is less parallel to the wall than that of Run Nos 2 and 4, respectively (the flow front is not shown here for Run Nos. 11 and 12). It is also seen that this prescribed distributions of concentration has little influence on the streamwise velocity distribution (see Fig.15). On the other hand, the influence on the transverse velocity distribution is clearly perceptible. While it has comparatively smaller influence on the transverse velocity when the concentration is low, the influence on the transverse velocity is much larger when the concentration is high. The low concentration near the walls has led to an increase in the transverse velocity.

Figure 16 show the effect of Reynolds number on the velocity distributions just behind the flow front at t=25. Here, the Reynolds number Re is taken as to be $Re_1$ as given by Eq.(15a). It is found that the decrease of Re reduces the transverse velocity, and makes the streamwise velocity changes very dramatically. From Fig.16(a), we observe that the low value of Re makes the centreline velocity become lower in magnitude than those regions on either sides of the centreline, and forms two "shoulders" with high magnitude of streamwise velocity thereby giving rise to a "W" like profile. It is further found by comparing Fig.12(a) and Fig.16(a) that this "shoulders" behaviour is generated either with the increase of concentration or the reduction of the Reynolds number. The Reynolds number influence on the velocity distribution just behind the front shown in Fig.16 is very interesting and this behaviour may have important impact on the product quality. As far as the knowledge of authors is concerned, this is for the first time that this Reynolds number effect on the velocity distribution behind the flow front is simulated at a given concentration of fibre suspension. More detailed studies for this behaviour would be very useful and this is the scope for future work.

**5.8 Further discussions**

The interaction between fibre orientation and the flow is further discussed below based on the simulated results. The influence of fibre on the flow focuses on the shear layer near the wall where the large first normal stress difference is generated. This normal stress difference will change the flow behaviour with the increase of concentration. For the Newtonian flow in a duct, the velocity distribution is parabolic far from the flow front. For the fibre suspension flow in a duct, the velocity distribution is flattened due to the effect of normal stress which increases toward the wall. Further, the streamwise velocity near the wall increases with the increase of concentration. In the fountain flow, with the increase of concentration, the increase of streamwise velocity near the wall will lead to the reduction of the transverse velocity owing to the mass conservation. Thus, the extent of fountain flow is lightened. As the result, the streamwise velocity in the core area is reduced. Therefore, in the region of the fountain flow, the streamwise velocity increases near the wall and decreases in the core area (Fig.12). At lower Re number, the roles of fibre stresses becomes more dominating. As such, in the region of the fountain flow, a velocity profile with "W" shape is produced with the increase of the fibre concentration (Fig.16a).

The reduction of Re diminishes the role of advection and the fibre stress plays an increasing role on the velocity distribution. The role of normal stress as produced by the fibre motion increases with the decrease of Re, which results in an anti-transversal motion of the fluid particles near the wall (role of fibre is largely manifested in the region near the wall) and reduces the fountain flow. Thus, the result with reducing Re number is that fibre stress leads to a streamwise velocity increase in the area near the wall. Correspondingly, the streamwise velocity in the central core area is reduced owing to the conservation of the total mass flow. This behaviour can also be found in Reddy and Mitchell (2001) in the duct for a contraction flow. The Fig.9 in this work showed that increasing the fibre stress reduces the streamwise velocity in the central core area.

For the shape of the front in a duct flow, Rose (1961) argued that the fluid interface will continuously have a constant curvature throughout when the fluid moves in a capillary channel of constant cross-section at constant flow rate. This description is based on the observation that a parabolic profile of axial velocity exists at all distant positions ahead of and behind the advancing interface. For the flow continuity, since a fountain effect exists behind the interface (in the side of wetting fluid), there must be an inverse fountain effect ahead of the interface (in the side of non-wetting fluid). At the interface, the tangential velocity must be equal from the both sides and thus it must be zero; it appears that the fountain and inverse fountain effects will cause opposite tangential velocities along the interface. Hence, the tangential pressure gradient along the interface is zero too in both sides. As such, the pressure difference across the interface along the interface is constant. According to the equilibrium of radial forces and surface tension, the curvature of the interface everywhere is constant owing to constant normal pressure difference. It seems that this statement is true from the simulation results in present study, since the shape of the flow front for Newtonian case is nearly a semi-circle at a large evolving time. This agreement also confirms that the numerical method and the simulation results in present study are correct.

In the presence of fibre suspension, the behaviour of the flow is no longer just purely Newtonian. The axial velocity deviates from the distribution of parabolic profile, as discussed before. The interface of the two immersible fluids is influenced by the fibre orientations, besides the surface tension and pressure gradient. Since some part of the pressure is balanced by the fibre stress in the side of wetting fluid, the pressure difference across the interface along the interface length is not constant any more. In the region where the fibre stress is high near the front, the change of the curvature of the front will be large. Therefore, the curvature of the interface varies along the interface length, as seen from the simulated results in this study.

The fountain flow has been also studied extensively for polymer melts in Tadmor (1974). The molecular orientation in injection molding was investigated theoretically and also assisted to some extent by experimental observations. The analysis contributes to a better understanding of the orientation distribution, its mechanism of formation, and its dependence on material properties and operating conditions. This work showed that the behaviour of shear flow (near the walls) behind the front is the source of the molecular orientation. The steady elongational flow, which is suggested to occur in the advancing front, explains the high values and the direction of orientation in the skin layer of the molding. The said work also indicates that the heat transfer in the process of injection molding is important for the final molecular orientation. These findings are in agreement with our simulations. The present study confirms that there is increasing elongational flow near the front (near the centerline) with the increase of concentration. The shear flow near the wall behind the front generates strong fibre orientation and large first normal stress difference with the increase of concentration.

## 6. Conclusions

The numerical method for the simulation of fibre orientation in suspension flow with front moving has been developed with the level set scheme and the projection method. The fibre contributions to the velocity field are accounted for by an ensemble configuration field method. The description of fibre motion with Jeffery equation allows the fibre motion to be modelled with the three roles of convection, rotation and diffusion. With the time evolution, the fibre stress is oscillating in time and space for limited number of configuration field. When the number of configuration field is large enough, this number has little influence on the final simulation result. The effects of various fibre parameters on the fibre orientation behaviour, stress distribution, velocity field, and front shapes are studied.

The effects of concentration and aspect ratio of fibres are set to alter the fibre stresses, and their influences on the flow front and velocity are different. The presence of fibres motion behaviour has little influence on the front shape in the ranges of fibre parameters studied at the fixed Reynolds number although this has large influence on the velocity. The effect of changing fibre parameters only gives rise to variation of front shape in the region near the wall whereas the front shape in the central core area is hardly affected. However, the fibre motion has strong influence on the distributions of the streamwise and transverse velocities in the fountain flow. Fibre motion produces a strong normal stress near the wall which leads to the large gradient of the stress near the wall which makes the streamwise velocity near the wall to increase; it also leads to a reduction of the transversal velocity on comparison to the pure Newtonian flow without fibres. Thus, the fibre addition to the flow weakens the strength of the fountain flow.

The effect of surface tension has a (relatively) large influence on the front shape. It has little or no influence on the velocity distribution and the fibre orientation far behind the front. This is perhaps not surprising as the surface tension is defined and only affects the flow around the front.

The fibre orientation is not always along the direction of velocity vector in the process of mold filling. The fibre is nearly aligned with the streamlines in the far upstream region of the flow front and near the walls. Along the centreline and in the region of fountain flow, the fibre orientation is almost crossing or traversing the velocity vector at large angle.

The effect of Re number ($Re_1$) on the velocity distribution and the fibre orientation is large. The decrease of Re reduces the transverse velocity, and makes the streamwise velocity changes dramatically. Decreasing the Reynolds number $Re_1$ for a given fibre concentration makes the magnitude of the streamwise velocity at centreline to become lower than those regions on either sides of the centreline, thereby leading to the formation of two "shoulders" with high magnitude of streamwise velocity. This finding is very interesting and it may have important impact on the product quality. This is perhaps for the first time that this Reynolds number effect on the velocity distribution is simulated at a given concentration of fibres.

Finally, as shown in this paper, the front flow evolution in fibre suspension flows can be simulated with the projection scheme and the level set method. With this algorithm, the effect of various material parameters on the flow properties, front shapes as well as the final fibre orientation can be obtained. This study will help to improve the product quality in injection moulding and other process industries.

**Acknowledgements**

The authors are grateful to X-J Fan (The University of Sydney) and Peter Kennedy (Moldflow Pty. Ltd) for their helpful discussions. They also wish to thank the anonymous reviewers for their helpful comments. Finally, the authors wish to thank the financial support from A*STAR and Moldflow company for the project (Project No. EMT/00/011).

**References**

Advani SG, Tucker CL (1987) The use of tensors to describe and predict fibre orientation in short fibre composites. J Rheol 31:751-784

Azaiez J, Guenette R (2002) Numerical Modelling of the Flow of Fibre Suspensions through a Planar Contraction. Canadian J of Chem Eng 80:1115-1125

Baloch A, Webster MF (1995) A computer-simulation of complex flows of fiber suspensions. Comput Fluids 24:135-151

Bell JB, Marcus DL (1992) A second-order projection method for variable density flows. J Comput Phys 101:334-348

Chang YC, Hou TY, Merriman B, Osher S (1996) A level set formulation of Eulerian interface capturing methods for incompressible fluid flows. J Comput Phys 124:449–464

Chiba K, Nakamura K (1990) A numerical solution for the flow of dilute fibre suspensions through an axisymmetric contraction. J Non-Newtonian Fluid Mech 35:1-14

Chiba K, Yasuda K, Nakamura K (2001) Numerical solution of fiber suspension flow through a parallel plate channel by coupling flow field with fiber orientation distribution. J Non-Newtonian Fluid Mech 99:145-157

Chinesta F, Poitou A (2002) Numerical analysis of the coupling between the flow kinematics and the fiber orientation in Eulerian simulations of dilute short fiber suspensions flows. Canadian J of Chem Eng 80:1107-1114

Dinh SH, Armstrong RC (1984) A rheological equation of state for semiconcentrated fiber suspensions. J Rheol 28:207-227

Dou HS, Phan-Thien N, Khoo BC, Yeo KS, Zheng R (2004) Simulation of front evolving liquid film flowing down an inclined plate using level set method. Comput Mech 34:271-281

Fan XJ, Phan-Thien N, Zheng Z (1999) Simulation of fibre suspension flows by the Brownian configuration field method, J. Non-Newt. Fluid Mech., 84, 1999, 257-274

Folgar FP, Tucker CL (1984) Orientation behavior of fibers in concentrated suspensions. J Reinforced Plastics and Composites. 3:98-119

Harlen OG, Koch DL (1997) Orientational drift of a fibre suspended in a dilute polymer solution during oscillatory shear f low. J Non-Newtonian Fluid Mech 73:81-97

Hinch EJ, Leal LG (1976) Constitutive equations in suspension mechanics. Part 2. Approximate forms for a suspension of rigid particles affected by Brownian rotations. J Fluid Mech 76:187-208

Hulsen MA, van Heel APG, van den Brule BHAA (1997) Simulation of viscoelastic flows using Brownian configuration fields. J Non-Newtonian Fluid Mech 70:79-101

Jeffery GB (1922) The motion of ellipsoidal particles immersed in viscous Fluid. Proc Roy Soc Lond A 102:161-179

Lipscomb II GG, Denn MM, Hur DU, Boger DV (1988) The flow of fiber suspensions in complex geometry. J Non-Newtonian Fluid Mech 26:297-325

Nguyen-Chung T, Menig G (2001) Non-isothermal transient flow and molecular orientation during injection mild filling. Rheol Acta 40:67-73

Ottinger HC, van den Brule BHAA, Hulsen MA (1997) Brownian configuration fields and variance reduced CONNFFESSIT. J Non-Newtonain Fluid Mech 70:255-261

Petrie CJS (1999) The rheology of fibre suspensions. J Non-Newtonian Fluid Mech 87:369-402

Phan-Thien N, Fan XJ, Tanner RI, Zheng R (2002) Folgar-Tucker constant for a fibre suspension in a Newtonian fluid. J Non-Newtonian Fluid Mech 103:251-260

Phan-Thien N, Graham AL (1991) A new constitutive model for fibre suspensions: flow past a sphere. Rheol Acta 30:44-57

Reddy BD, Mitchell GP (2001) Finite element analysis of fibre suspension flows. Comput Method Appl Mech & Eng 190:2349-2367

Rose W (1961) Fluid-fluid interfaces in steady motion. Nature 191:242-243

Rosenberg J, Denn M (1990) Simulation of non-Recirculating flows of dilute fibre suspensions. J Non-Newtonian Fluid Mech 37:317-345

Sethian JA, Smereka P (2003) Level set methods for fluid interfaces. Annu Rev Fluid Mech 35:341-372

Shaqfeh ESG, Fredrickson GH (1990) The hydrodynamic stress in a suspension of rods. Phys Fluids A 2:7-24

Sussman M, Fatemi E, Smereka P, Osher SJ (1998) An improved level set method of incompressible two-fluid flows. Comput Fluids 27:663-680

Sussman M, Smereka P, Osher SJ (1994) A level set approach to computing solutions to incompressible two-phase flow. J Comput Phys 114:146–59

Tadmor Z (1974) Molecular orientation in injection molding. J Applied Polymer Sci 18:1753-1772

VerWeyst BE, Tucker CL (2002) Fiber Suspensions in Complex Geometries: Flow/Orientation Coupling. Canadian J of Chem Eng 80:1093-1106

Yasuda K, Kyuto T, Mori N (2004) An experimental study of flow-induced fiber orientation and concentration distributions in a concentrated suspension flow through a slit channel containing a cylinder. Rheol Acta 43:137-145

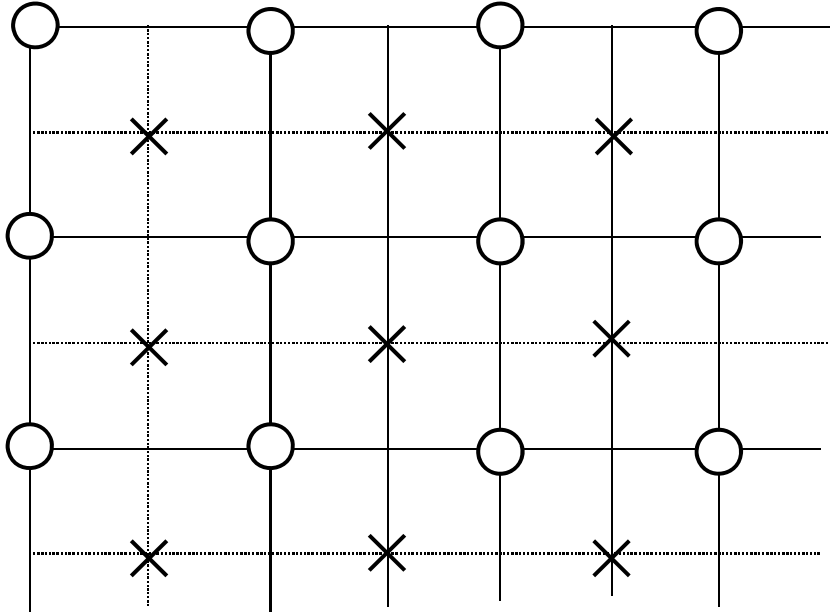

Fig. 1 Schematic of the grid arrangement. Primary grid points O: for $\mathbf{u}$, $\rho$, $\mu$, $\phi$; Dual grid points ×: for $\psi_t$, $\nabla\mathbf{u}$, $\mathbf{q}$, $\boldsymbol{\tau}$, and the resulting $p$

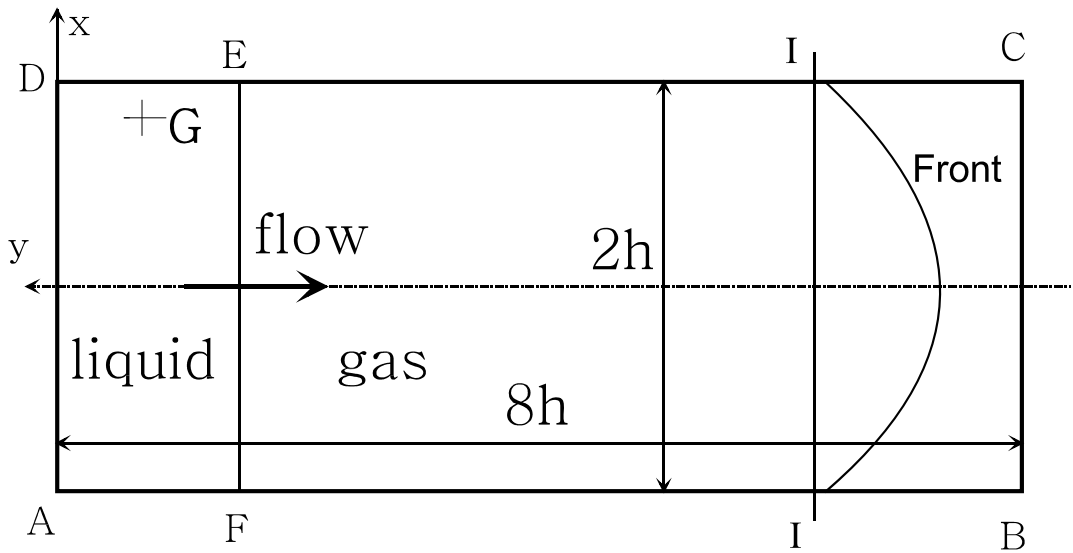

Fig. 2 Schematic of the computing domain and coordinates arrangement

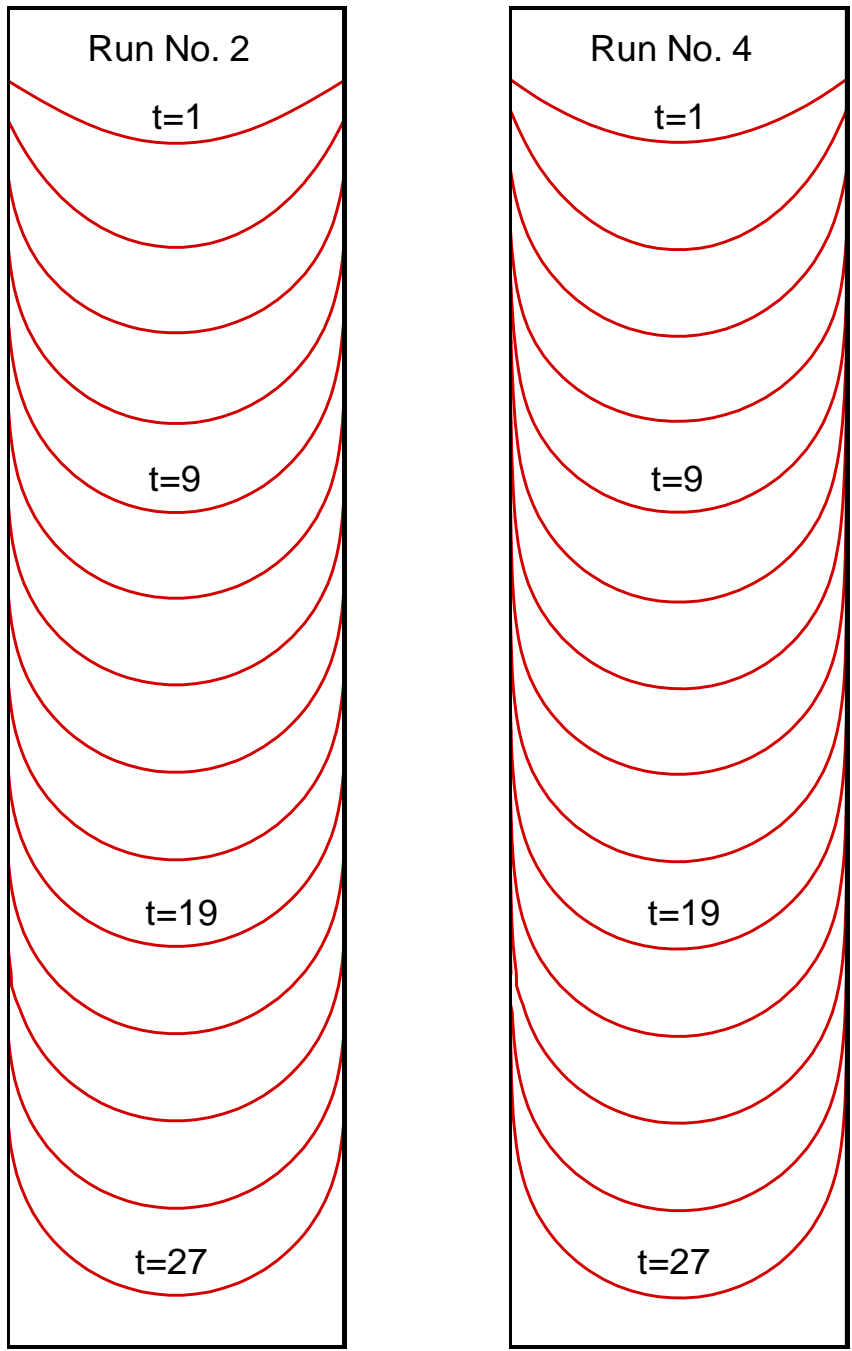

Fig. 3 Computing results for the flow front evolution with the time in the fibre suspension flow. Left: $\varphi = 0.10$, $a_r = 10$, $N = 1000$, $C_i = 0.01$, K=5.20, and We=3.044; Right: $\varphi = 0.25$, $a_r = 10$, $N = 500$, $C_i = 0.01$, K=40.86 and We=3.044. The time interval between two contour lines is 200 $\Delta t$

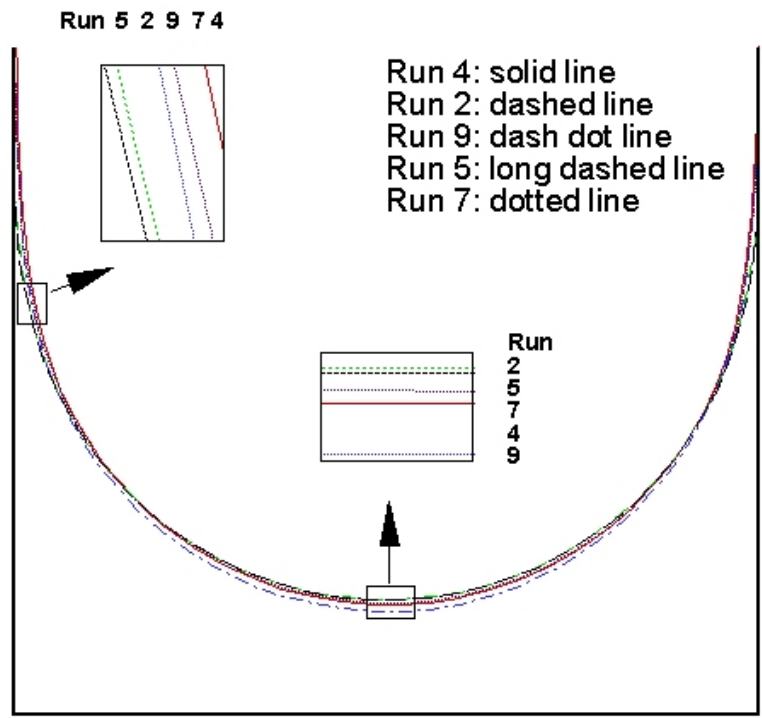

Fig. 4 Computing results for the flow front evolution with the time in the fibre suspension flow at t=27. Run No.5 is the Newtonian fluid case

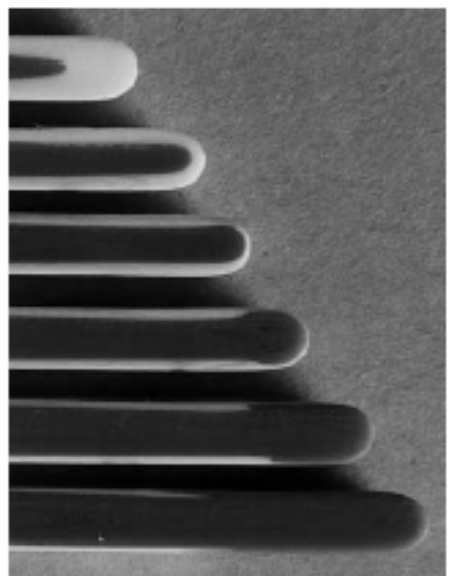

Fig. 5 Flow fronts and isochrones in sandwich injection molded short shots at different times (Nguyen-Chung and Menig, 2001)

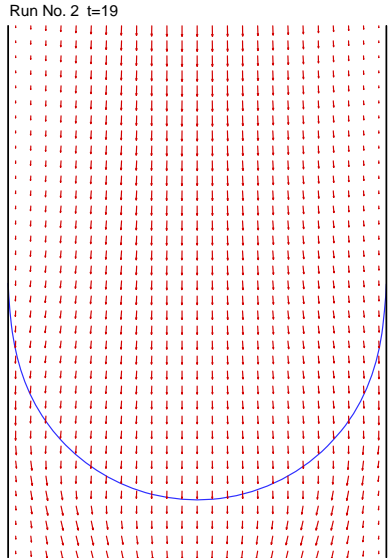 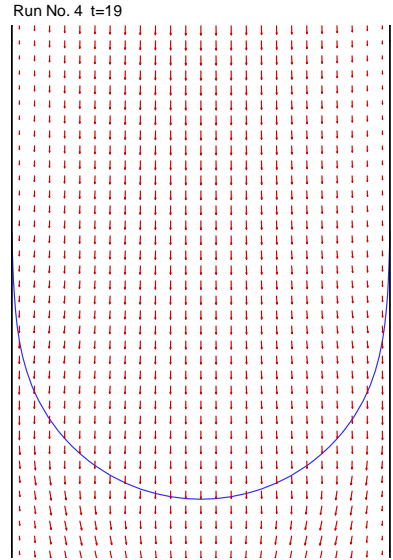

(a)                 (b)

Fig. 6 Computing results for the velocity vector around the flow front at t=19 for various sets of parameters. (a) $\varphi = 0.10$, $a_r = 10$, $N = 1000$, $C_i = 0.01$, K=5.20, and We=3.044. (b) $\varphi = 0.25$, $a_r = 10$, $N = 1000$, $C_i = 0.01$, K=40.86 and We=3.044

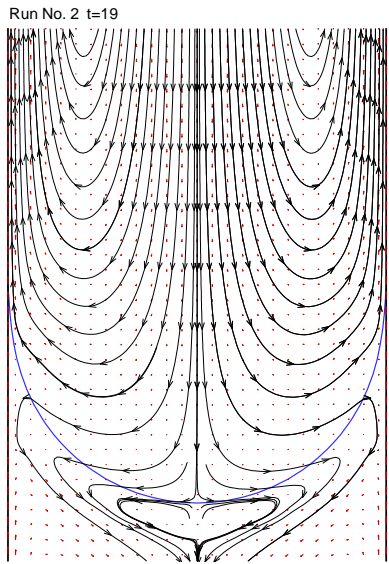 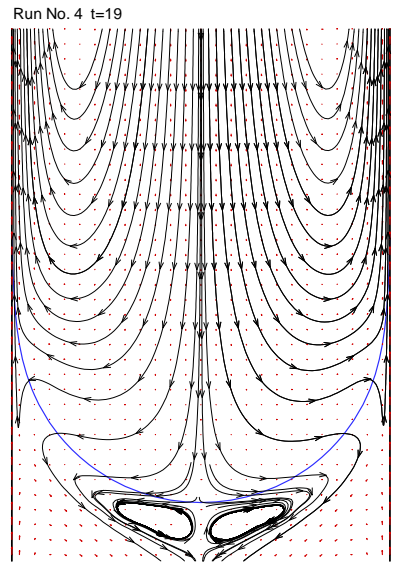

(a)                 (b)

Fig. 7 Streamlines with the moving coordinates fixed at the fore front at t=19 for various sets of parameters. (a) $\varphi = 0.10$, $a_r = 10$, $N = 1000$, $C_i = 0.01$, K=5.20, and We=3.044. (b) $\varphi = 0.25$, $a_r = 10$, $N = 1000$, $C_i = 0.01$, K=40.86 and We=3.044

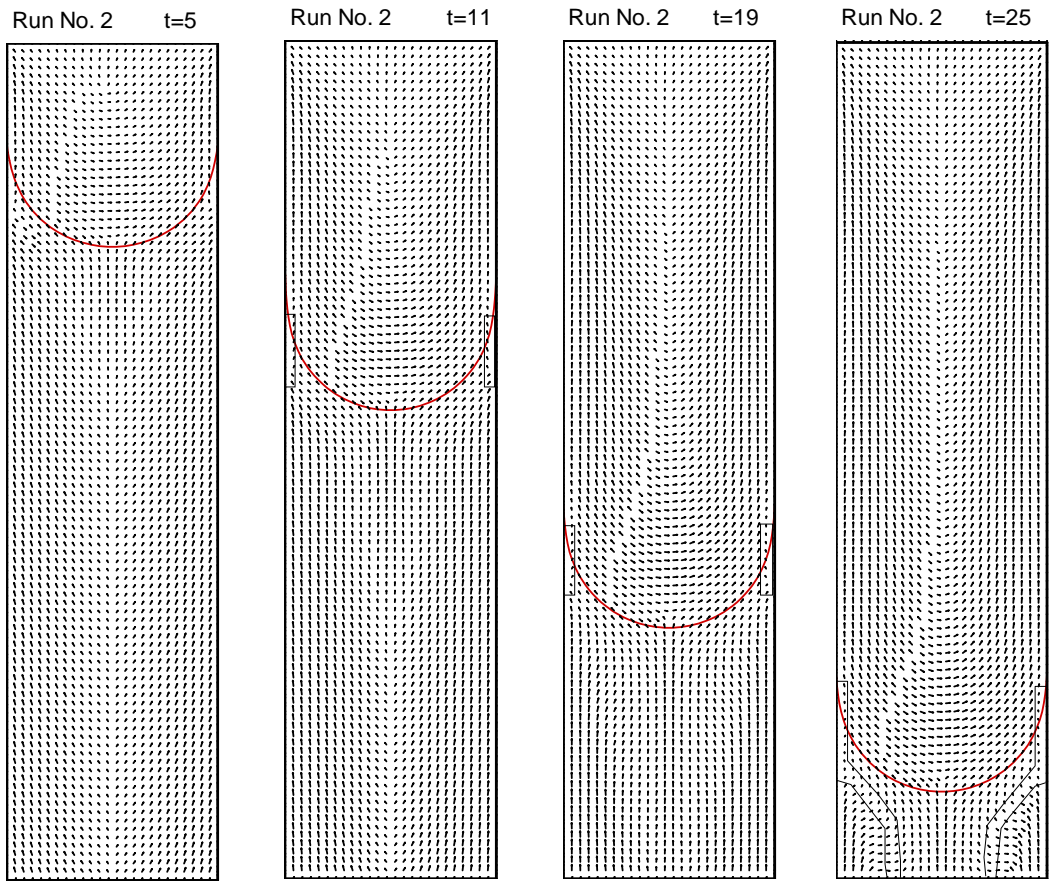

Fig. 8 Simulation results of fibre orientation with the time evolution in the suspension flow. The fibre parameters: $\varphi = 0.10$, $a_r = 10$, $N = 1000$, $C_i = 0.01$, K=5.20, and We=3.044

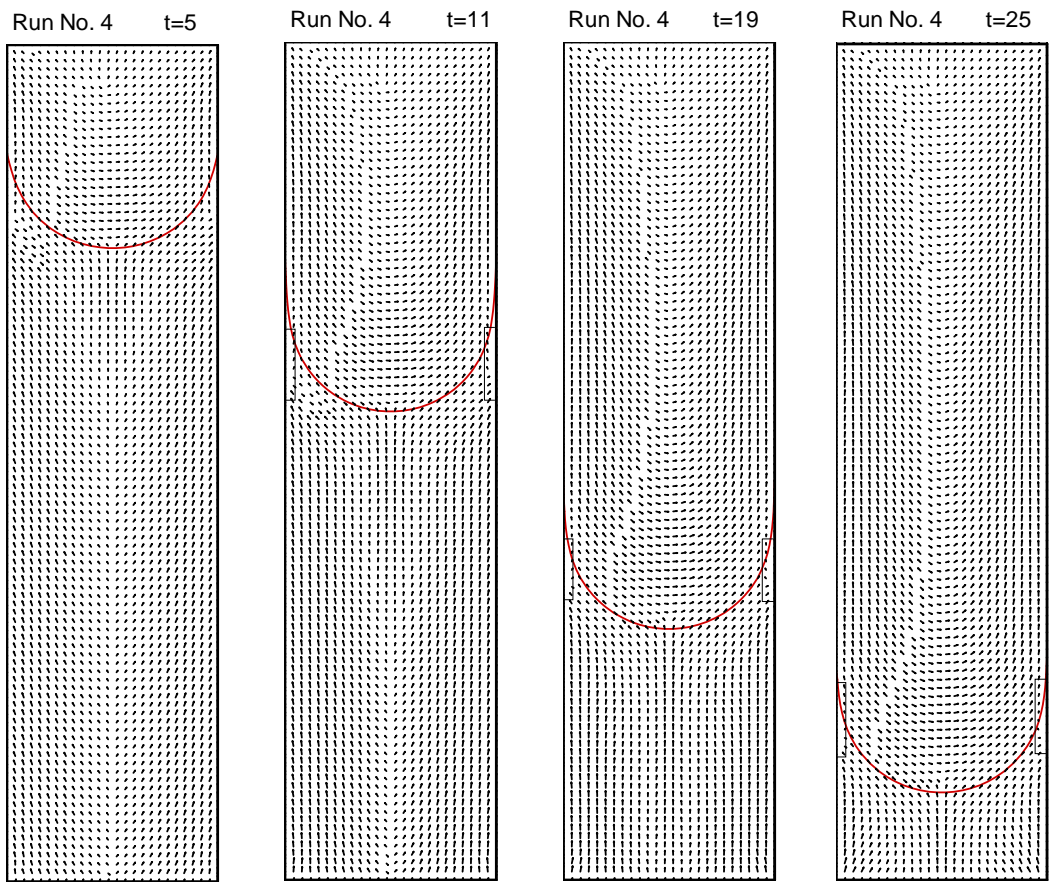

Fig. 9 Simulation results of fibre orientation with the time evolution in the suspension flow. The fibre parameters: $\varphi = 0.25$, $a_r = 10$, $N = 1000$, $C_i = 0.01$, K=40.86 and We=3.044

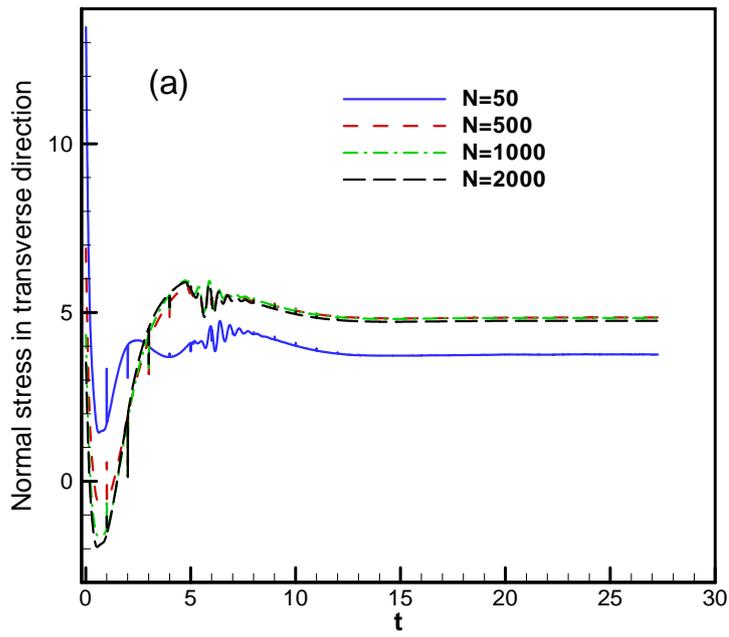

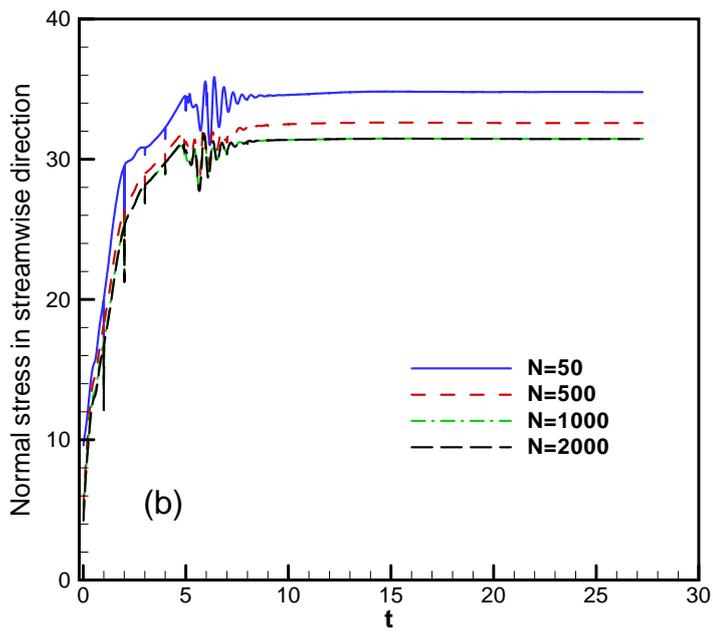

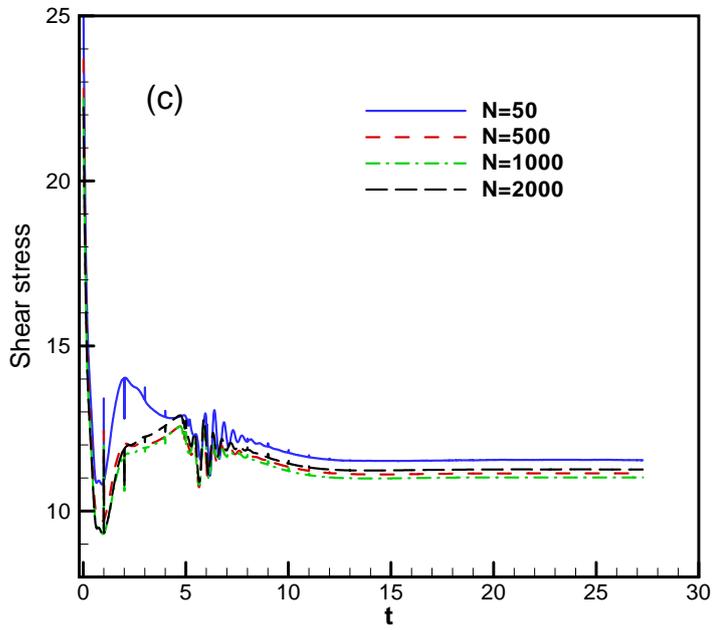

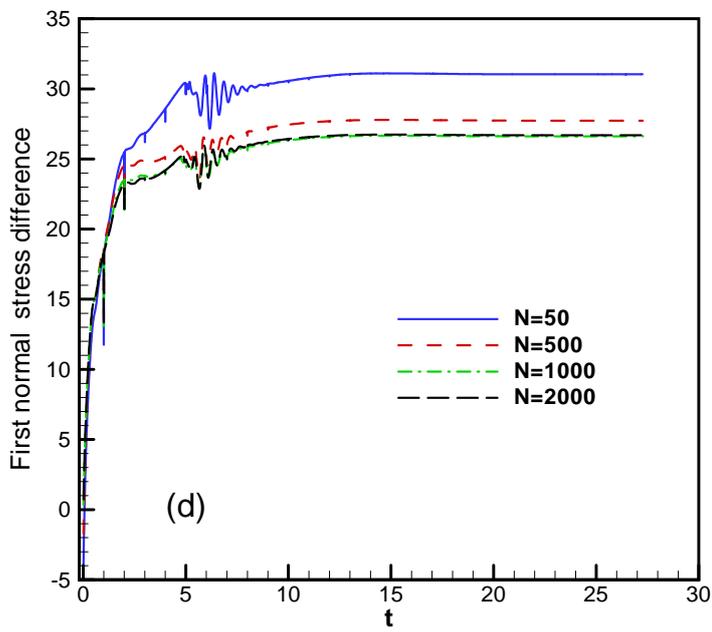

Fig. 10 Effect of the configuration number on the calculations for the stresses with the time (Run No.4). (a) The normal stress $\tau_{f_{xx}}$; (b) The normal stress $\tau_{f_{yy}}$; (c) The shear stress $\tau_{f_{xy}}$; (d) The first normal stress difference $N_1 = \tau_{f_{yy}} - \tau_{f_{xx}}$. The values of these stresses are recorded at the position E as shown in Fig.1. The fibre parameters: $\varphi = 0.25$, $a_r = 10$, $C_i = 0.01$, K=40.86 and We=3.044

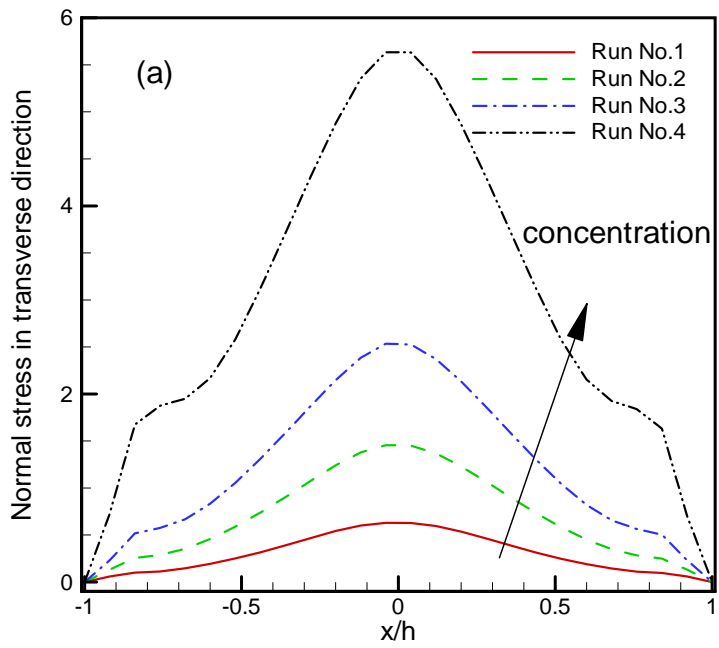

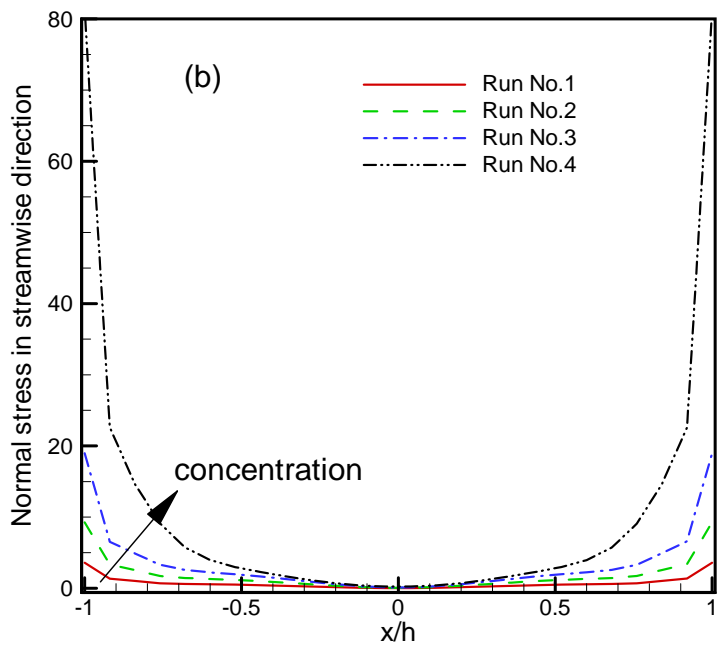

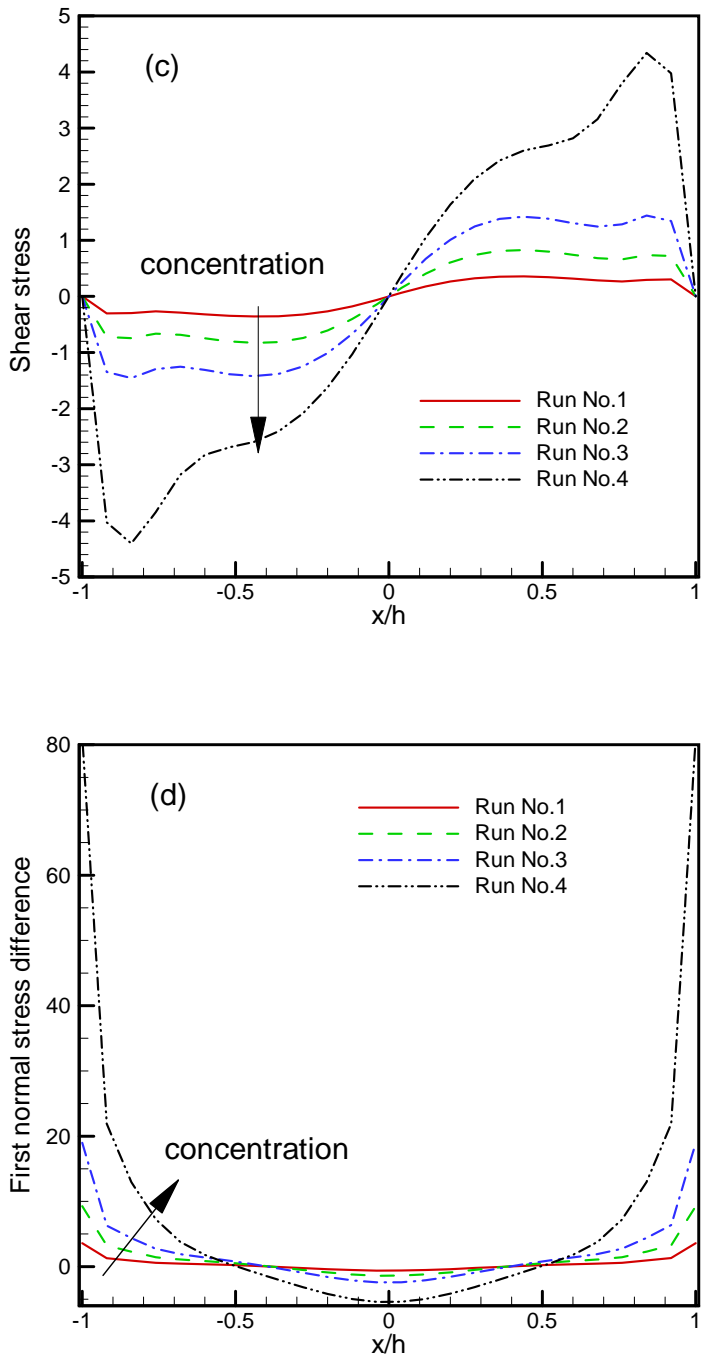

Fig. 11 Effects of concentration on the stress distributions along the channel width at the position just behind the front at t=25. The normal stress $\tau_{f_{xx}}$; (b) The normal stress $\tau_{f_{yy}}$; (c) The shear stress $\tau_{f_{xy}}$; (d) The first normal stress difference $N_1 = \tau_{f_{yy}} - \tau_{f_{xx}}$. The concentration is $\varphi = 0.05$, 0.10, 0.15, and 0.25 respectively. The other fibre parameters are $a_r = 10$, $N = 1000$, $C_i = 0.01$, and We=3.044

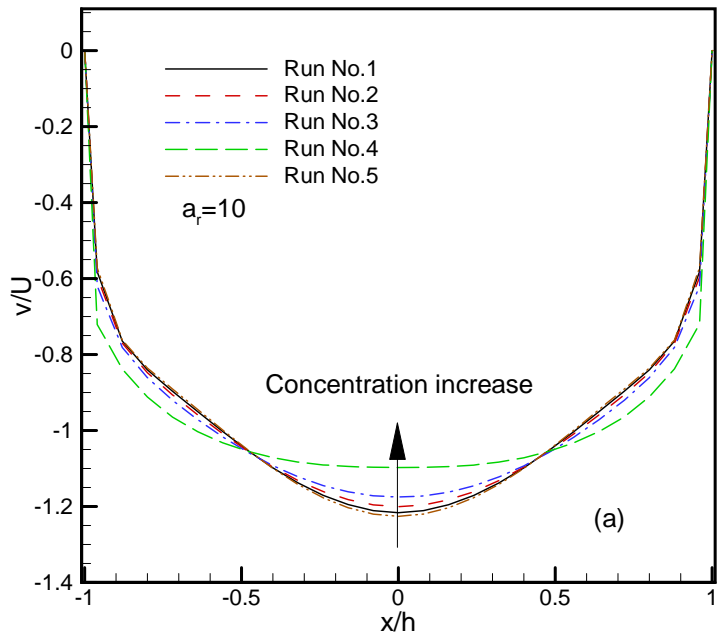

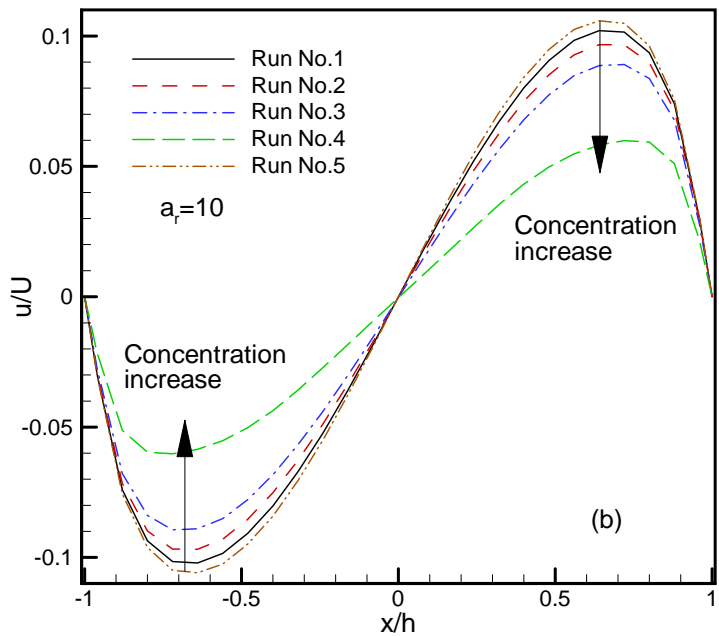

Fig. 12 Effects of concentration on the velocity distributions at the position just behind the front at t=25. (a) Streamwise velocity profile. (b) Transverse velocity profile. The concentration is $\varphi = 0.05$, 0.10, 0.15, and 0.25 respectively. The other fibre parameters are $a_r = 10$, $N = 1000$, $C_i = 0.01$, and We=3.044

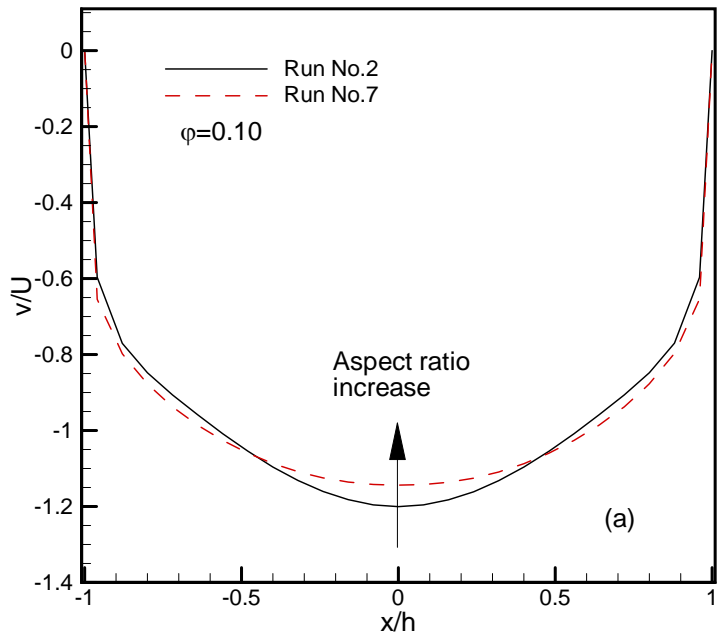

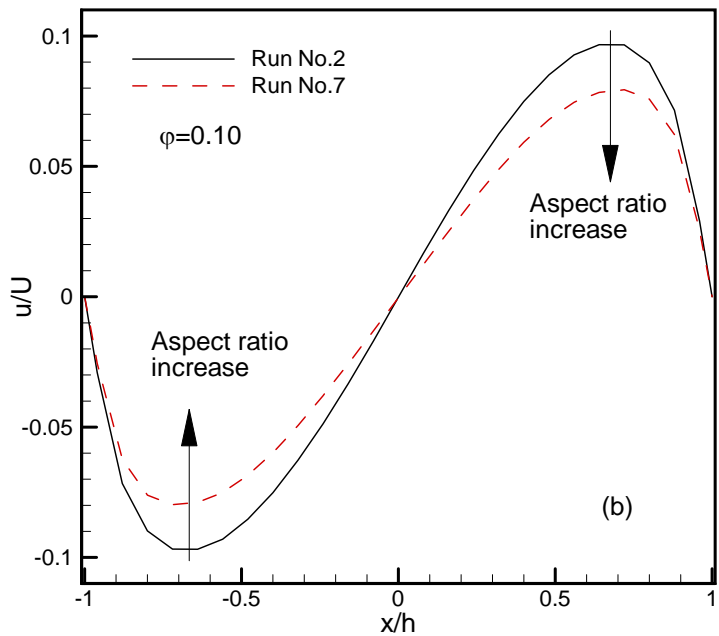

Fig. 13 Effects of aspect ratio of fibres on the velocity distributions at the position just behind the front at t=25. The aspect ratio is $a_r = 10$ and 20 respectively. (a) Streamwise velocity profile. (b) Transversal velocity profile. The other fibre parameters are $\varphi = 0.10$, $N = 1000$, $C_i = 0.01$, and We=3.044

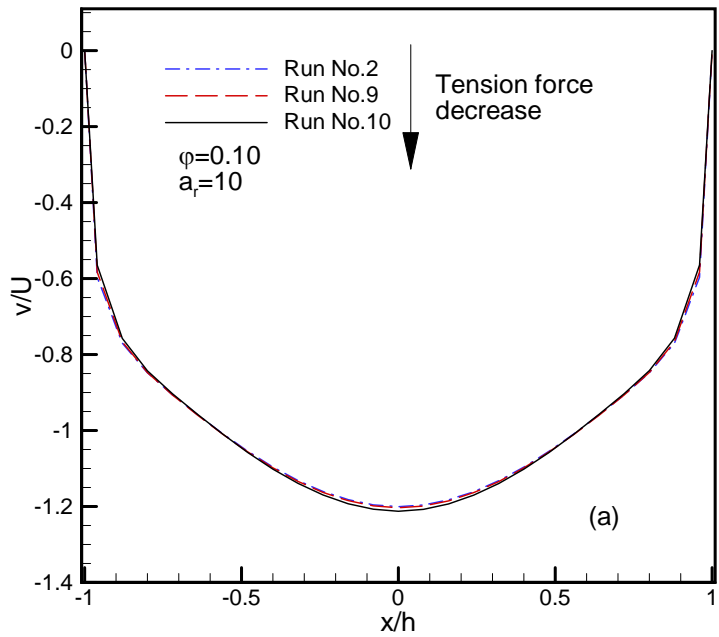

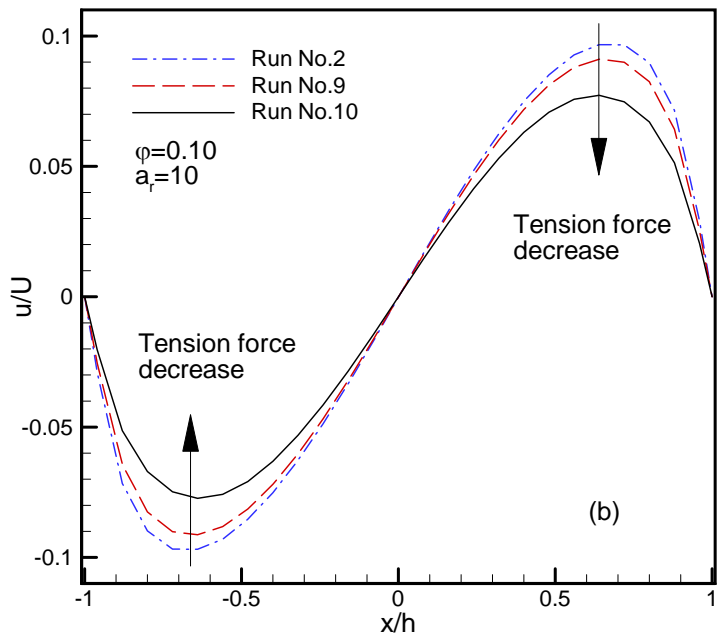

Fig. 14 Effects of surface tension coefficient (or Weber number) on the velocity distributions at the position just behind the front at t=25. (a) Streamwise velocity profile. (b) Transversal velocity profile. The other fibre parameters are $N=1000$ and $C_i=0.01$. The Weber number is We=3.044, We=5.555, and We=11.11

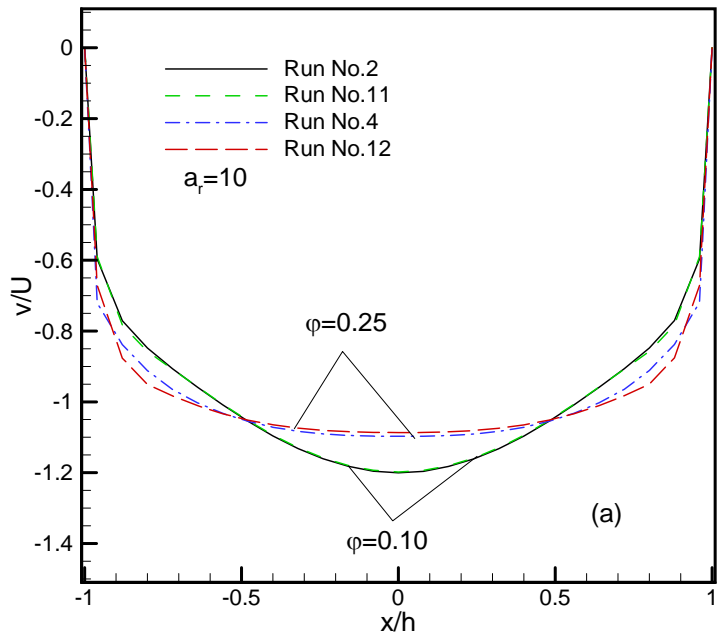

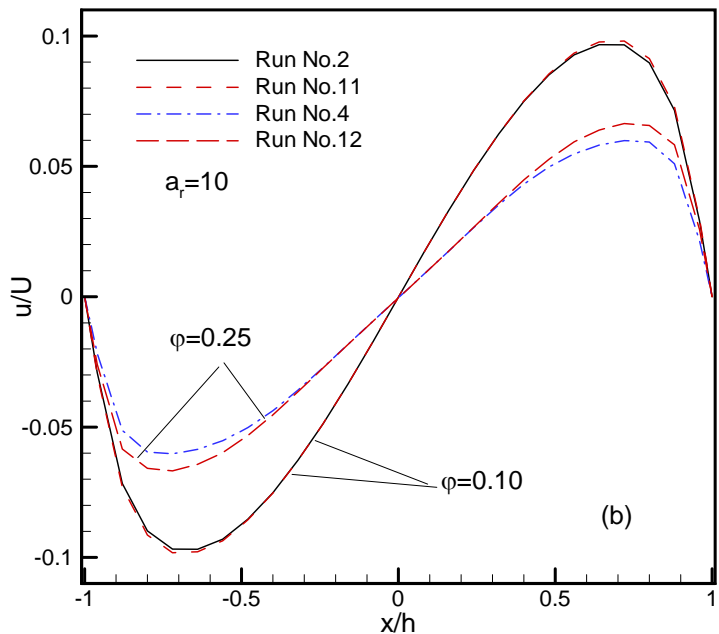

Fig. 15 Effects of distribution of concentration on the velocity distributions at the position just behind the front at t=25. (a) Streamwise velocity profile. (b) Transversal velocity profile. The other fibre parameters are $N=1000$, $C_i=0.01$, and We=3.044. Run No. 2 and 4 stand for the uniform distribution of the concentration. Run No.11 and 12 represent that with prescribed non-uniform distributions of the concentration in Eq.(43)

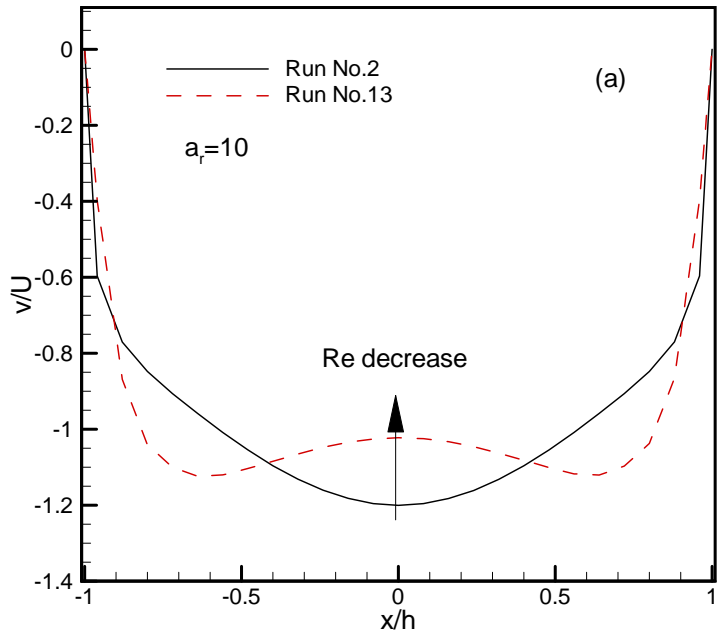

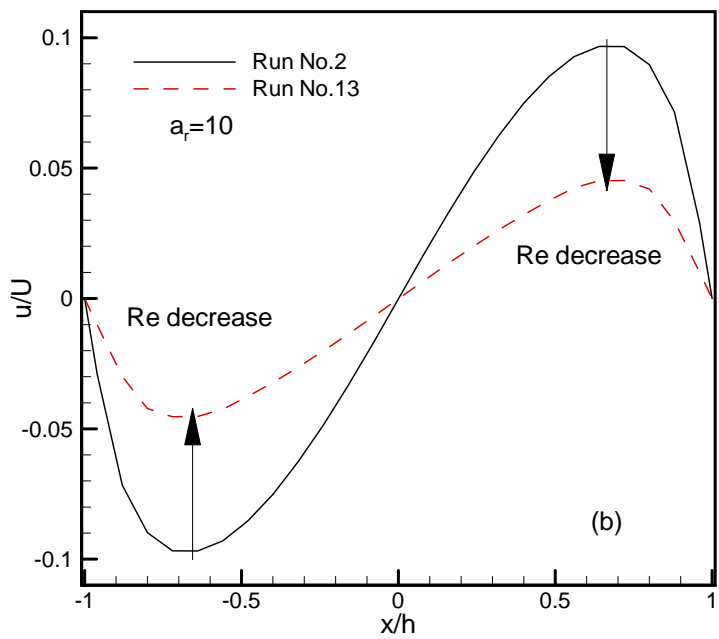

Fig. 16 Effects of Reynolds number on the velocity distributions at the position just behind the front at t=25 for $Re_1$=300 and $Re_1$=167. (a) Streamwise velocity profile. (b) Transversal velocity profile. The other fibre parameters are $a_r = 10$, $\varphi = 0.10$, $N = 1000$, $C_i = 0.01$, and We=3.044